\documentclass{article}

\PassOptionsToPackage{numbers}{natbib}

\usepackage[preprint]{neurips_2025}

\usepackage[utf8]{inputenc} 
\usepackage[T1]{fontenc}    
\usepackage{hyperref}       
\usepackage{url}            
\usepackage{booktabs}       
\usepackage{amsfonts}       
\usepackage{nicefrac}       
\usepackage{microtype}      
\usepackage{xcolor}         

\usepackage{neurips_2025}
\usepackage{multirow}
\usepackage[ruled]{algorithm2e}

\usepackage{amsmath,amsfonts,bm}

\def\eqref#1{equation~(\ref{#1})}

\def\1{\bm{1}}

\def\vh{{\bm{h}}}

\def\vq{{\bm{q}}}

\def\vu{{\bm{u}}}
\def\vv{{\bm{v}}}
\def\vw{{\bm{w}}}
\def\vx{{\bm{x}}}
\def\vy{{\bm{y}}}
\def\vz{{\bm{z}}}

\def\mK{{\bm{K}}}

\def\mM{{\bm{M}}}

\def\mR{{\bm{R}}}

\def\mT{{\bm{T}}}

\def\mV{{\bm{V}}}

\DeclareMathAlphabet{\mathsfit}{\encodingdefault}{\sfdefault}{m}{sl}
\SetMathAlphabet{\mathsfit}{bold}{\encodingdefault}{\sfdefault}{bx}{n}

\usepackage{graphicx} 
\usepackage{float} 
\usepackage{booktabs}
\usepackage{tabularx}
\usepackage{subcaption}
\usepackage{caption}
\usepackage{wrapfig}

\title{MeshCoder: LLM-Powered Structured Mesh Code Generation from Point Clouds}

\usepackage{authblk}
\usepackage{textcomp}

\author[1,2]{\textbf{Bingquan Dai}$^{*}$}
\author[1]{\textbf{Li Ray Luo}$^{*}$}
\author[1,3]{\textbf{Qihong Tang}}
\author[1,4]{\textbf{Jie Wang}}
\author[1]{\textbf{Xinyu Lian}}
\author[1,5]{\textbf{Hao Xu}}
\author[2]{\textbf{Minghan Qin}}
\author[1]{\textbf{Xudong Xu}}
\author[1]{\textbf{Bo Dai}}
\author[2]{\textbf{Haoqian Wang}$^{\dagger}$}
\author[1]{\textbf{Zhaoyang Lyu}$^{\dagger}$}
\author[1]{\textbf{Jiangmiao Pang}}

\affil[1]{Shanghai Artificial Intelligence Laboratory, Shanghai, China}
\affil[2]{Tsinghua University, Beijing, China}
\affil[3]{Harbin Institute of Technology, Shenzhen, China}
\affil[4]{Beijing Institute of Technology, Beijing, China}
\affil[5]{AI Thrust, HKUST(GZ), Guangzhou, China}

\date{}  

\begin{document}

\maketitle

\let\thefootnote\relax
\footnotetext{$^{*}$Equal contribution.}
\footnotetext{$^{\dagger}$Corresponding authors.}

\vspace{-3em}
\begin{figure}[htbp]
  \centering
  \begin{minipage}[b]{\textwidth}
    \includegraphics[width=\textwidth]{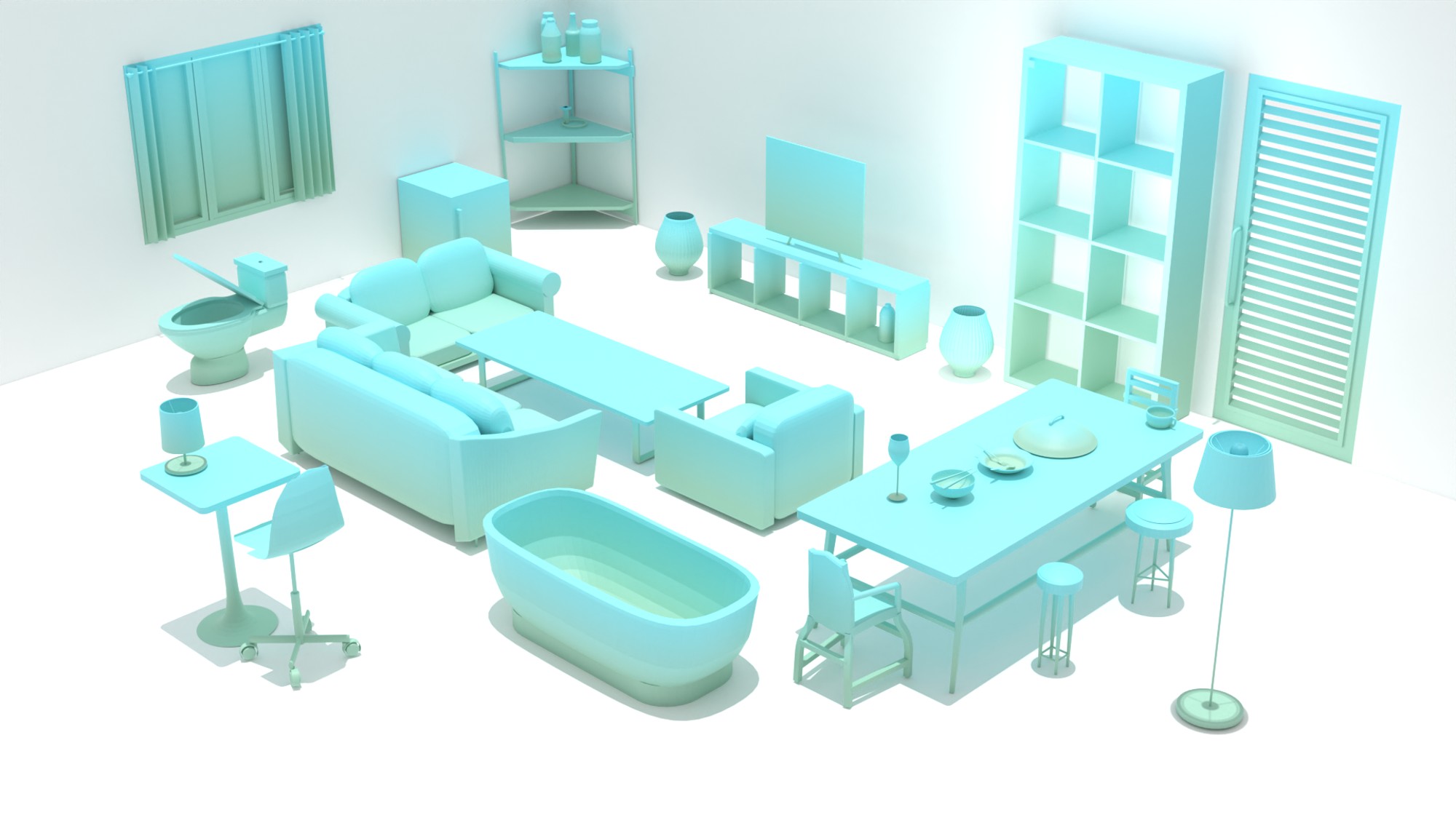}
    \vspace{-3em}
    \subcaption{} 
  \end{minipage}
  \vspace{-0.5em}
  \begin{minipage}[b]{\textwidth}
    \includegraphics[width=\textwidth]{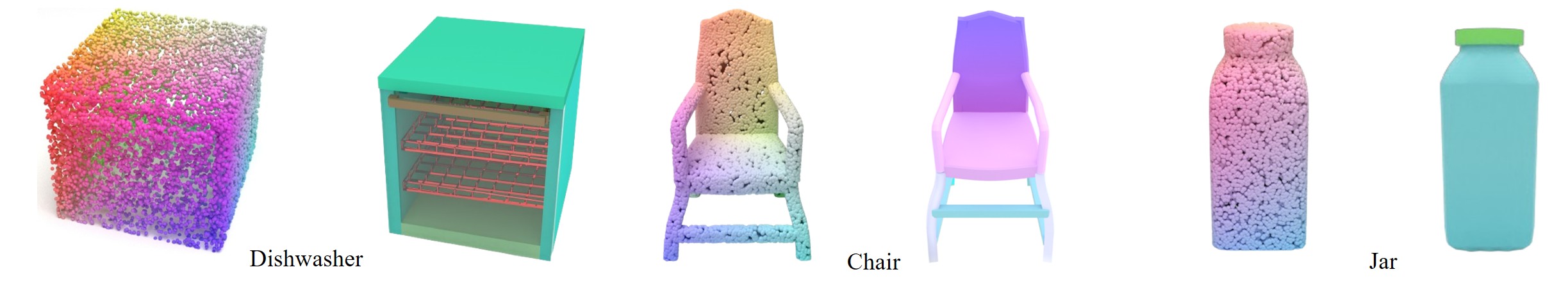}
    \subcaption{}
  \end{minipage}
  \caption{ (a) MeshCoder can predict codes and reconstruct 41 categories of objects. 
  (b) MeshCoder takes in point clouds and produce part-segmented meshes by executing the predicted code in Blender.
  For the dishwasher, we apply transparency to the foremost part to showcase the internal structure.}
  \label{fig:minipage-example}
\end{figure}
\begin{abstract}

Reconstructing 3D objects into editable programs is pivotal for applications like reverse engineering and shape editing. However, existing methods often rely on limited domain-specific languages (DSLs) and small-scale datasets, restricting their ability to model complex geometries and structures. To address these challenges, we introduce MeshCoder, a novel framework that reconstructs complex 3D objects from point clouds into editable Blender Python scripts. We develop a comprehensive set of expressive Blender Python APIs capable of synthesizing intricate geometries. Leveraging these APIs, we construct a large-scale paired object-code dataset, where the code for each object is decomposed into distinct semantic parts. Subsequently, we train a multimodal large language model (LLM) that translates 3D point cloud into executable Blender Python scripts. Our approach not only achieves superior performance in shape-to-code reconstruction tasks but also facilitates intuitive geometric and topological editing through convenient code modifications. Furthermore, our code-based representation enhances the reasoning capabilities of LLMs in 3D shape understanding tasks. Together, these contributions establish MeshCoder as a powerful and flexible solution for programmatic 3D shape reconstruction and understanding. Project homepage: \url{https://daibingquan.github.io/MeshCoder}.

\end{abstract}
\vspace{-0.2cm}
\section{Introduction}
\label{sec:intro}


Inferring shape programs from 3D observations is
of great importance for reverse engineering, shape editing, and 3D structure understanding.
Prior work~\cite{tian2018learning, jones2022plad, liang2022learning} has explored this problem by defining Domain-Specific Languages (DSLs) to model geometric and structural properties of objects and training neural networks to map 3D observations to shape programs.
However, existing methods struggle to generalize to objects with complex geometry and structure. Two key limitations underlie this gap. 
First, existing DSLs are constrained to modeling simple primitives (e.g., cubes, spheres, cylinders) and cannot represent real-world objects with intricate parts.
Second, training shape-to-code inference models demands large-scale paired datasets of 3D objects and their corresponding code, while such datasets are scarce.
Prior work often relies on datasets with limited categories, geometric complexity and part count.

To address these challenges, we introduce MeshCoder, a novel framework for generating Blender Python scripts that reconstruct complex 3D objects into their constituent parts.
First, we design a set of expressive Blender Python APIs that are capable of synthesizing intricate geometries beyond simple primitives. For instance, our APIs can create complex shapes by translating a 2D section curve along a specified trajectory, bridging section curves of different shapes, adding bevels or applying Boolean operations on basic shapes, repeating a basic shape in one dimension or two dimensions.
With these concise yet powerful Blender Python APIs, 
we can model highly complex shapes, addressing the limitations of prior DSLs.

Second, we present a novel pipeline to construct a large-scale paired object-code dataset. We begin by synthesizing diverse object parts using our APIs with parametrically sampled parameters, yielding a part-level dataset. A part-to-code inference model is then trained on this dataset to predict code for individual parts.
Next, we employ this model to construct a holistic object-code dataset.
We use Infinigen-Indoor~\cite{infinigen2024indoors} to generate a dataset of objects, and each object is decomposed into its constituent parts.
We use the part-to-code inference model to predict code for each part of an object, and then carefully design rules to concatenate code of all parts to obtain code of the object.
This process yields a dataset of approximately 1 million objects spanning 41 categories, with objects up to more than 100 parts.
Finally, we train a multimodal large language model (LLM) on this dataset to infer code from 3D objects.
We use point clouds as 3D shape representations due to their ease of acquisition, and use a triplane-based tokenizer to transform the input point cloud to a set of fixed-length tokens.
These tokens are fed into the LLM to generate Blender Python scripts that replicate input geometries in distinct semantic parts.

We evaluate our approach against existing shape-to-code methods, with experimental results and quantitative metrics demonstrating that our framework significantly outperforms prior work. Furthermore, by representing shapes as executable code, our method facilitates intuitive geometric and topological editing through simple code modifications. This capability enables precise alterations to object geometry and mesh topology, enhancing flexibility in downstream applications. Additionally, we conduct experiments on shape structural and geometric understanding tasks, revealing that our code-based representation improves the reasoning capabilities of large language models (LLMs) when interpreting 3D shapes.
In summary, our contributions are outlined as follows:
\begin{itemize}
    \item We have developed a comprehensive set of Blender Python APIs, facilitating the modeling of intricate geometries. This enhanced API suite empowers the procedural generation of complex 3D structures, effectively addressing the limitations of traditional domain-specific languages (DSLs) in representing detailed and varied shapes.
    \item We propose a pipeline to construct a large-scale paired object-code dataset. Using the dataset we constructed, we can train an shape-to-code inference model. 
    \item We trained MeshCoder, an \textbf{Object-to-Code inference framework} that generates Blender Python scripts to reconstruct 3D meshes from point clouds in a structured and editable manner. Our model encodes 3D shapes into part-level code, simplifying mesh editing and enhancing LLMs' understanding of 3D objects.
\end{itemize}

\begin{figure}[ht]
    \centering
    \includegraphics[width=14cm]{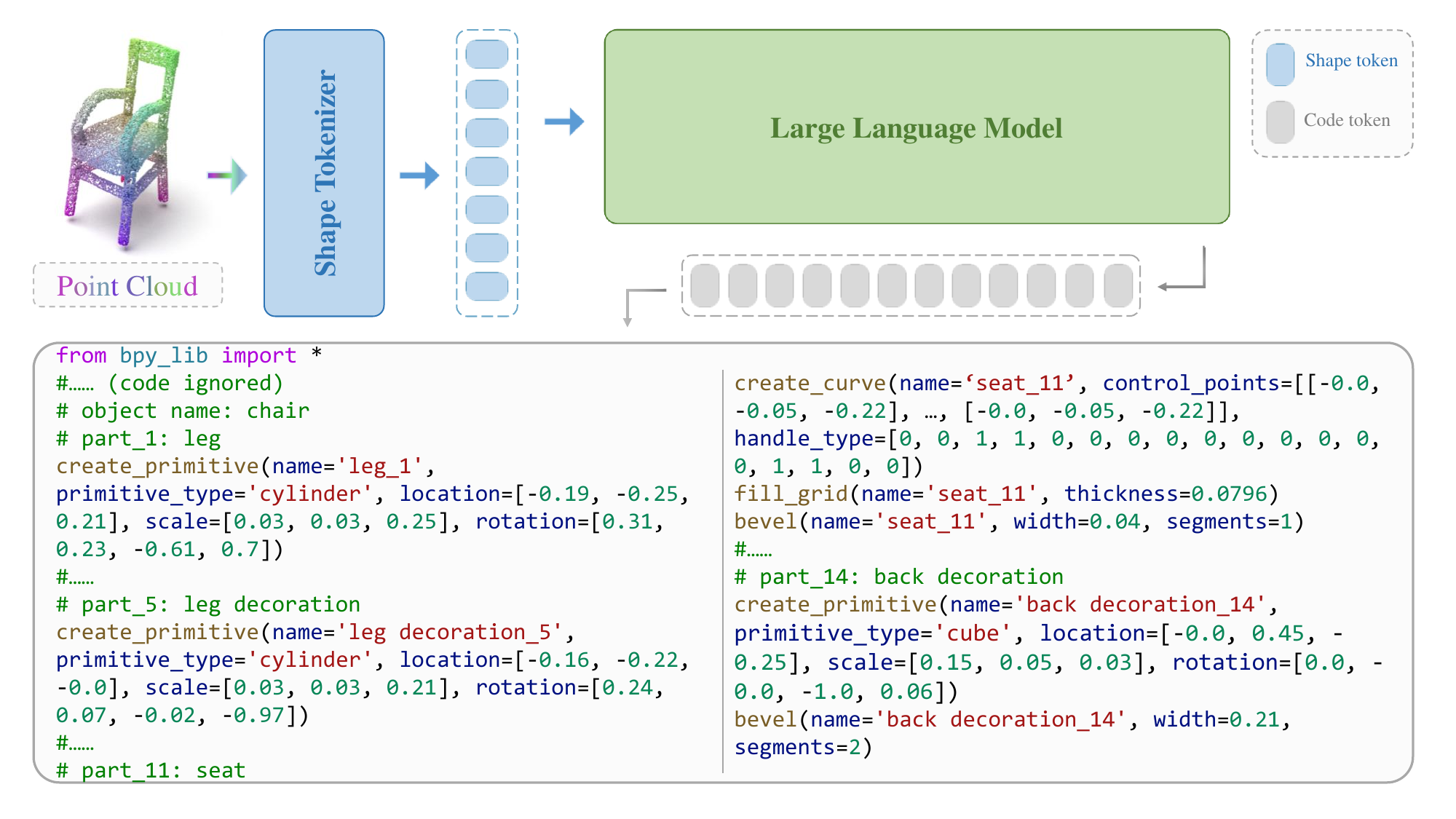}
    \vspace{-10pt}
    \caption{Overview of MeshCoder. The input point cloud is first encoded into shape tokens via a shape tokenizer. These tokens are then fed into a large language model (LLM), which autoregressively generates executable code representing part-based 3D structures. The decoded code specifies object's name, part identities and names, enabling interpretable and modular reconstruction.}
    \label{fig:pipeline}
    \vspace{-10pt}
\end{figure}
\section{Related Work}
\label{sec:related_work}

\textbf{Shape programs.}  
Shape programs provide a structured and interpretable framework for representing 3D geometry by utilizing domain-specific languages to describe the generative processes of shapes. Early work such as ShapeAssembly~\cite{jones2020shapeassembly} introduced explicit shape programs that capture the hierarchical and part-based organization of objects. Subsequent methods, including ShapeCoder~\cite{jones2023shapecoder}, PLAD~\cite{jones2022plad}, and ShapeLib~\cite{jones2025shapelib}, progressively improved program abstraction, learning efficiency, and scalability with large language models. Other approaches, such as those proposed by Liang~\cite{liang2022learning} and Tian et al.~\cite{tian2018learning}, incorporate differentiable rendering or neuro-symbolic reasoning to enhance program inference and execution. While these methods exhibit strong generalization capabilities in composing simple geometric elements like boxes and cylinders, they often struggle to model complex part geometries or generate artist-grade quad meshes, which restricts their application in high-fidelity asset creation. In addition, a range of methods in CAD program generation~\cite{Jayaraman2022SolidGenAA,Khan_2024_CVPR,9710909,xu2024CADMLLM,xu2023hierarchical,2504.20830} have explored synthesizing code representations for individual CAD parts. However, these approaches are limited to isolated component generation and lack the capability to model complete multi-part objects with coherent structural relationships.

\paragraph{Part-based Representation.}
Part-based representations have proven highly valuable in 3D shape analysis and synthesis. Some approaches~\cite{koo2023salad, liu2024part123, chen2024partgen, huang2024part, gao2019sdm, wu2019sagnet, mo2019structurenet, wu2020pq, nakayama2023difffacto, petrov2023anise} take a generative approach, assembling objects by combining predefined or learned parts into complete 3D structures.
Other methods~\cite{yang2024sampart3d, zhao2021point, abdelreheem2023satr, wang2019voxsegnet, liu2023partslip, zhou2023partslip++, xue2023zerops, umam2024partdistill, tang2024segment, thai20243, zhong2024meshsegmenter, zhu2023pointclip} focus on segmenting 3D objects into individual parts, enabling more modular and flexible manipulation of shapes. For instance, SAMPart3D~\cite{yang2024sampart3d} introduces a scalable zero-shot 3D part segmentation framework that segments any 3D object into semantic parts at multiple granularities without requiring predefined part label sets as text prompts. PartSLIP~\cite{liu2023partslip} explores low-shot part segmentation of 3D point clouds by leveraging a pretrained image-language model, GLIP, transferring rich knowledge from 2D to 3D through GLIP-based part detection on point cloud rendering and a novel 2D-to-3D label lifting algorithm. SATR~\cite{abdelreheem2023satr} performs zero-shot 3D shape segmentation via text descriptions by using a zero-shot 2D object detector, inferring 3D segmentation from multi-view 2D bounding box predictions by exploiting the topological properties of the underlying surface.
Despite these advancements in part segmentation and reconstruction, these methods do not translate segmented parts into executable code representations, limiting their integration into code-driven design workflows.
\section{Methodology}
\label{sec:method}
As shown in Figure~\ref{fig:pipeline}, we aim to train an object-to-code inference model that takes in a point cloud of an object, and then predict the Blender python scripts of each part of the object.
When executing the python scripts in Blender, we can obtain the same object in separated parts. 
To train such an object-to-code inference model, we need a dataset of paired objects and the corresponding codes.
To obtain such a dataset, we first train a part-to-code inference model that predicts code for a single part on our synthetic dataset of paired parts and the corresponding codes.
Then, given a dataset of objects separated in different parts, 
we use the trained part-to-code inference model to predict code for every part of an object.
Finally, we concatenate the codes of every part of the object and obtain the code of the object.
Now, we have a dataset of paired objects and the corresponding codes, and are ready to train the object-to-code inference model.

We explain the key steps described above in details in the following sections.
First, we explain how to synthesize a dataset of paired parts and the corresponding codes in Section~\ref{sec: part_and_code_dataset}.
Then, we describe the training procedure of the part-to-code inference model in Section~\ref{sec: Part-to-Code Reconstruction Model}.
Next, we use the part-to-code inference model to obtain the code of an entire object in Section~\ref{sec: Assemble Parts to Objects}.
Finally, we train the object-to-code inference model in Section~\ref{sec: Object-to-Code Reconstruction Model}.

\subsection{Part Dataset}
\label{sec: part_and_code_dataset}
\begin{figure}[ht]
    \vspace{-0.3cm}
    \centering
    \includegraphics[width=14cm]{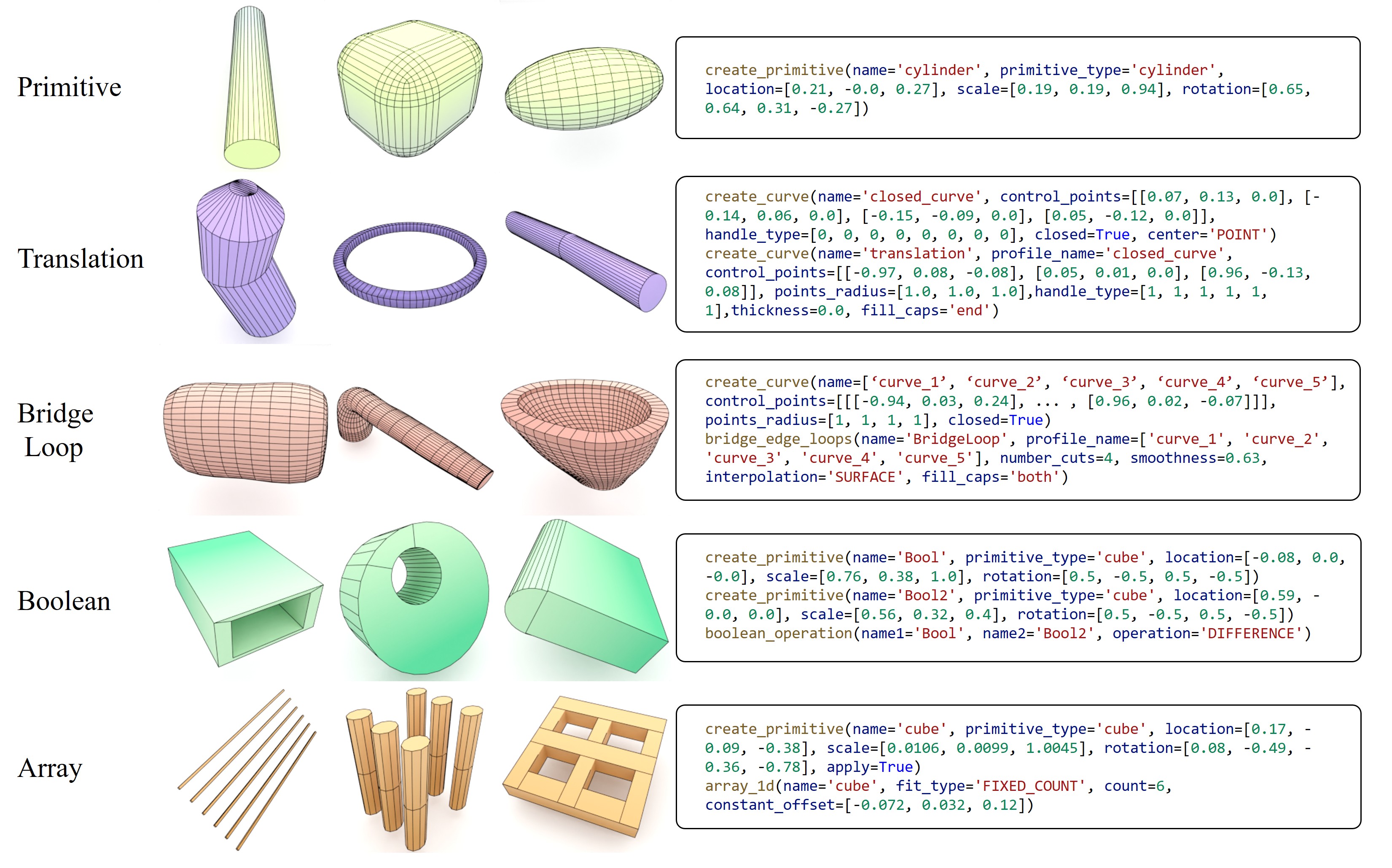}
    \caption{Visualization of basic geometric shape types and their corresponding code. For each shape category, the code shown corresponds to the first example.}
    \vspace{-10pt}
    \label{fig:all_parts}
\end{figure}

We aim to generate a dataset of paired part shapes and codes.
To do so, we implement probabilistic programs to generate Blender Python scripts, and obtain the corresponding shape by executing the code in Blender.
We carefully design these probabilistic programs and ensure that the shapes generated are within the range $[-1,1]^3$.
There are several types of shapes that we generate, as illustrated in Figure ~\ref{fig:all_parts}. We explain them in the following paragraphs.

\textbf{Primitive.}  
Primitives are a set of fundamental geometric shapes, consistent with those defined in Blender. Specifically, we consider five basic shapes: \texttt{cube}, \texttt{cylinder}, \texttt{UV sphere}, \texttt{cone}, and \texttt{torus}. Each primitive is parameterized by three attributes: \texttt{location} (\( \mathbf{location} \in \mathbb{R}^3 \)), \texttt{rotation} (\( \mathbf{rotation} \in \mathbb{H} \)), and \texttt{scale} (\( \mathbf{scale} \in \mathbb{R}^3 \)), where \( \mathbb{H} \) denotes the space of unit quaternions. The \texttt{location} specifies the shape's position in 3D space, \texttt{rotation} defines its orientation via quaternions, and \texttt{scale} determines the shape's size along its local axes. Examples of Primitives can be found in the first row of Figure ~\ref{fig:all_parts}.

\textbf{Translation.}
Translation is defined as the geometry obtained by sweeping a 2D cross-sectional shape along a 3D trajectory curve. As illustrated in the second row of Figure ~\ref{fig:all_parts}, during this translation process, the tangent direction of the 3D trajectory remains perpendicular to the 2D shape, and the size of the section shape can change along the 3D trajectory.For a more detailed explanation, please refer to \ref{appendix: datasets}. To implement this, we first define a 2D shape using a set of control points (i.e., spatial coordinates), and then specify a 3D trajectory curve in a similar manner. Specifically, our experiments consider five types of cross-sectional shapes: rectangles, circles, circular arcs, polygons, and Bézier curves. 
For the trajectories, we define six forms: straight lines, polylines, circles, circular arcs, rectangles, and Bézier curves. 
Notably, this method also allows a 2D shape to rotate around an axis to form a solid of revolution, making it suitable for modeling objects such as bottles and plates.

\textbf{Bridge loop.} 
Although the Translation method is capable of generating certain complex objects, it remains constrained in several ways. For instance, in the Translation operation, the 2D cross-sectional shape is always orthogonal to the tangent direction of the trajectory. Moreover, the section shape is allowed to change only in scale, without any deformation in its geometry.
To address this limitation, we introduce an alternative method for constructing geometries, namely the Bridge Loop. This geometry is constructed by first generating a sequence of 2D shapes and then connecting their corresponding vertices to form a continuous 3D geometry. Some cases can be seen in the third row of Figure ~\ref{fig:all_parts}. The Bridge Loop approach enables the creation of more complex geometries compared to those achievable via Translation alone. For a more detailed explanation, please refer to \ref{appendix: datasets}.

\textbf{Boolean.}
Boolean geometries refer to geometries formed by applying Boolean operations—namely union, intersection, and difference—to two or more of the fundamental shape categories defined in Section~\ref{sec: part_and_code_dataset}.
The union operation enables the construction of complex composite geometries, while the difference operation is used to generate geometries with holes or indentations. We can see some examples and their corresponding codes in the fourth row of Figure ~\ref{fig:all_parts}.

\textbf{Array.}
When a particular type of primitive geometry appears repeatedly in a regular pattern, we do not invoke the construction function for each primitive individually, as this would result in lengthy code. 
Instead, we employ an Array method to construct the entire structure collectively. Specifically, we define two types of Arrays: 1D Arrays, where a geometry is repeatedly instantiated along a curve, and 2D Arrays, where repetition occurs across a plane. Cases of this type can be seen in the last row of Figure ~\ref{fig:all_parts}.

\vspace{-0.05cm}
\subsection{Part-to-code Inference Model}
\label{sec: Part-to-Code Reconstruction Model}
After constructing the dataset of paired code $\vy$ and mesh $\mM$, we sample a point cloud $\vx \in \mathbb{R}^{N \times 3}$ from each mesh $\mM$, where $N$ is the number of points in the point cloud. 
We train a part-to-code inference model $\vh$ that takes in a point cloud $\vx$ and predict the corresponding code $\vy$.
The inference model consists of two modules: The shape tokenizer model and a fintuned LLM.
The tokenizer model takes in the point cloud $\vx$ and outputs a set of fixed length tokens $\vz \in \mathbb{R}^{L \times D}$, where $L$ is the number of shape tokens and $D$ is the dimension of each token.
We set $D$ to the same dimension as the word embeddings in the LLM.
Thereafter, the LLM takes in the shape tokens $\vz$ and then predict $\vy$, the code of the point cloud $\vx$.
We train the shape tokenizer model and finetune the LLM at the same time using the cross-entropy loss for the prediction of the next token in the shape code $\vy$.
We use Llama-3.2-1B as the base LLM and finetune it using LoRA.
\paragraph{The shape tokenizer model.}
\begin{figure}[ht]
    \vspace{-0.5cm}
    \centering
    \includegraphics[width=14cm]{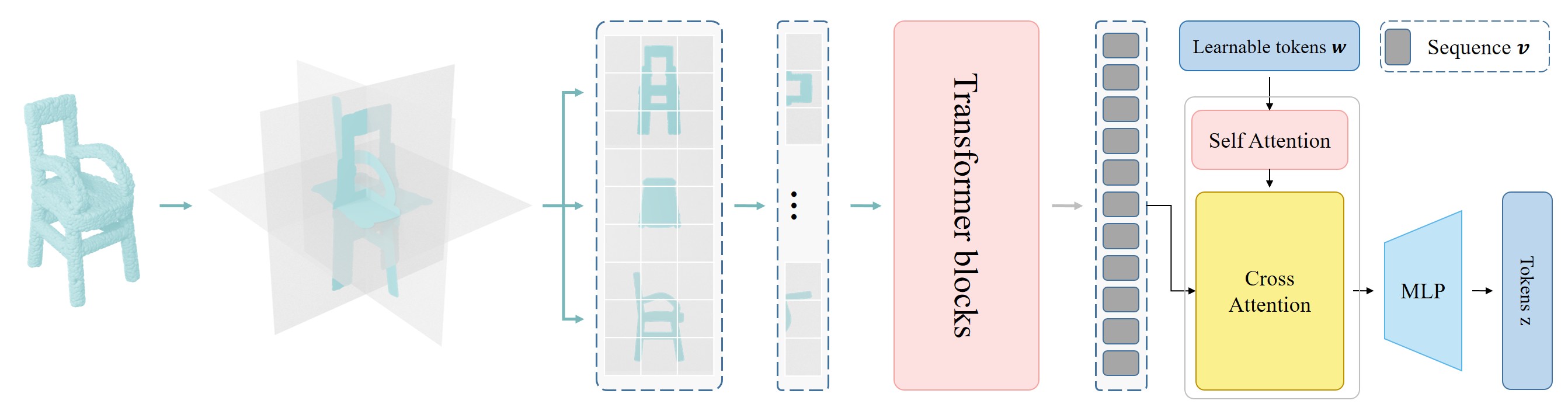}
    \caption{Architecture of the shape tokenizer. We first project the point cloud into the triplane and obtain triplane features.
    The triplane features are patchified and reshaped into a 1D sequence, and fed into transformer blocks to obtain triplane tokens. 
    Finally, we use a set of learnable tokens to aggregate information from triplane tokens via cross-attention.
    }
    \label{fig:shape_tokenizer}
    \vspace{-0.25cm}
\end{figure}

We explain the detailed structure of the shape tokenizer model.
As shown in Figure~\ref{fig:shape_tokenizer}, the shape tokenizer model transforms a point cloud $\vx \in \mathbb{R}^{N \times 3}$ to a set of fixed length tokens $\vz \in \mathbb{R}^{L \times D}$.
We first project the point cloud $\vx$ to a triplane and obtain triplane feature 
$\vu \in \mathbb{R}^{3 \times H \times W \times D_1}$, where $H, W$ are the height and width of the planes, and $D_1$ is the dimension of the triplane feature.
The coordinates of each point is fed to a shared MLP and a feature of dimension $D_1$ is obtained.
We project each point's feature to the three
perpendicular planes according to the point's position.
Features projected to the same pixel are aggregated by max-pooling.
Pixels that do not correspond to any point are filled with zeros.
After obtaining the triplane feature $\vu$, we patchify it and reshape it into a $1D$ sequence $\vv \in \mathbb{R}^{(3 \cdot H/f \cdot W/f) \times D_1}$, where $f$ is the patch size.
We then feed the sequence $\vv$ to a set of transformer blocks and outputs $\vv' \in \mathbb{R}^{(3 \cdot H/f \cdot W/f) \times D_1}$.
Next, to compress the number of tokens fed into the LLM, we use a learnable set of tokens $\vw \in \mathbb{R}^{L \times D_2}$ to aggregate information from $\vv'$ using cross attention:
\begin{align}
    \text{CrossAttn}(\text{Transformer}(\vw), \vv', \vv'),
\end{align}
where $\text{Transformer}$ denotes a transformer block, 
$\text{CrossAttn}(Q, K, V)$ denotes a cross attention block, and $Q, K, V$ are query, key, value, respectively.
By feeding $\vw$ to a set of these cross attention blocks, we obtain tokens $\vw' \in \mathbb{R}^{L \times D_2}$ that contain information about the point cloud $\vx$.
Finally, we use an MLP to transform the dimension of $\vw'$ from $D_2$ to $D$ and obtain shape tokens $\vz \in \mathbb{R}^{L \times D}$, where $D$ is the dimension of the word embeddings in the LLM.
Now, the shape tokens $\vz$ can be readily fed to the LLM and predict the code corresponding to the point cloud $\vx$.

\subsection{Assemble Parts to Objects}
\label{sec: Assemble Parts to Objects}
\begin{figure}[h]
    \vspace{-0.5cm}
    \centering
    \includegraphics[width=14cm]{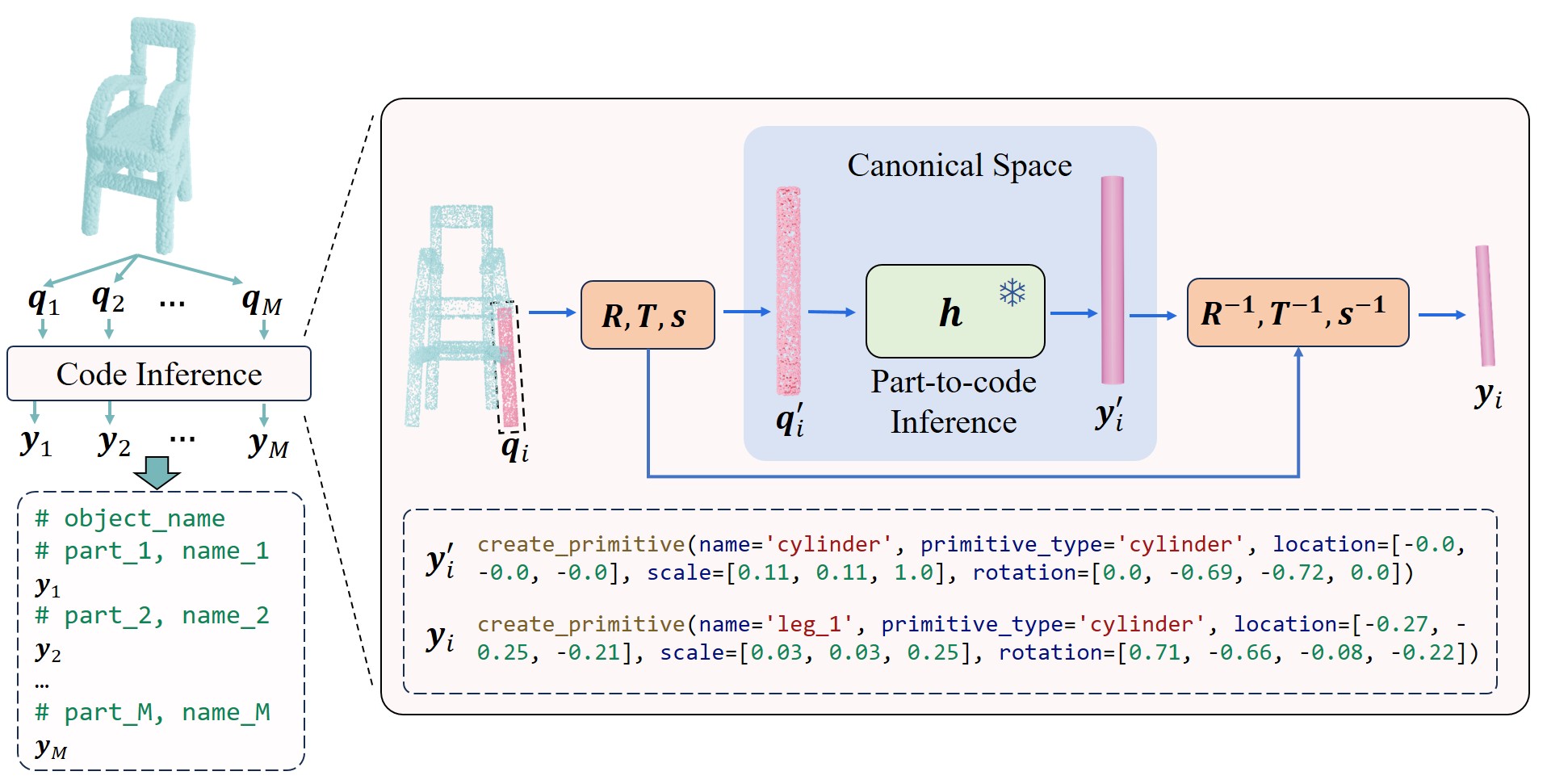}
    \vspace{-10pt}
    \caption{
    Pipeline of object-level code dataset construction using the part-to-code inference model. For each part point cloud $\vq_i$, the code inference module independently predicts its corresponding code $\vy_i$. All part codes $\vy_i$ are then concatenated to form the complete object code.
    We also add meaningful semantic information to the object code following the template shown in the figure.
    The complete code of the example chair is shown in Figure~\ref{fig:pipeline}.}
    \label{fig:code_concat}
    \vspace{-10pt}
\end{figure}

After training the part-to-code inference model $\vh$, we can use it to obtain the code of an object.
Given a dataset of objects, in which each object $\mathcal{O}$ is separated into its constituent parts $\mathcal{O} = \{\vq_i | i=1,2,\cdots,M\}$, where $\vq_i$ is the $i$-th part of object $\mathcal{O}$, and $M$ is the number of parts of the object $\mathcal{O}$.
We also assume that each part $\vq_i$ has its semantic label.
We can use the part-to-code inference model $\vh$ to obtain the code of each part.
Specifically, we first normalize each part $\vq_i$ to the cube $[-1,1]^3$ according to its minimum bounding box and obtain the shape $\vq_i'$.
Then we use the part-to-code inference model $\vh$ to obtain its code $\vy_i'=\vh(\vq_i')$.
We then implement algorithms to transform the relevant numerical parameters in the code $\vy_i'$ to the original location, scale, and pose of $\vq_i'$ and obtain the code $\vy_i$ of the original shape $\vq_i$.
Finally, we concatenate the codes of all parts of the object, add semantic information to the code for each part, and obtain the code of the object $\vy= \{\vy_i | i=1,2,\cdots,M\}$. When concatenating the code, we sort each part based on its spatial position. Specifically, we assign an index to each part following a spatial order from bottom to top, left to right, and front to back.
An overview of this pipeline is illustrated in Figure~\ref{fig:code_concat}.
During code inference, the part point cloud $\vq_i$ is first transformed into a canonical space using a rotation matrix $\mR$, translation matrix $\mT$, and scaling factor $s$, resulting in $\vq_i'$. 
The trained part-to-code inference model $\vh$ generates the code $\vy_i'$ of $\vq_i'$. 
$\vy_i'$ is then transformed back to the original pose and scale using the inverse of $\mR$, $\mT$, and $s$, and we obtain the code $\vy_i$ of $\vq_i$.

\subsection{Object-to-code Inference Model}
\label{sec: Object-to-Code Reconstruction Model}
After obtaining the code $\vy$ of each object $\mathcal{O}$ in the dataset, we can use them to train an object-to-code inference model.
Our object-to-code inference model has the same structure as the part-to-code inference model described in Section~\ref{sec: Part-to-Code Reconstruction Model}.
We initialize the weights of the object-to-code inference model as the weights of the trained part-to-code inference model, and use the same training method in Section~\ref{sec: Part-to-Code Reconstruction Model} to train the object-to-code inference model.
It is worth noting semantic information in the ground-truth code of objects enables the object-to-code inference model to learn the semantic structure of objects, and facilitate 3D shape understanding.

\section{Experiment}
\label{sec:experiment}

\subsection{Datasets}

\subsubsection{Synthetic Part Dataset}
To facilitate the training of our part-to-inference model, we first constructed a synthetic part dataset. Specifically, we utilized functions from our basic shape code library, randomly sampling their parameters based on manually defined distributions to generate paired data of synthetic parts and corresponding code. This process yielded 1.5 million point cloud–code pairs for primitive shapes, 3 million for Translation-based parts, 1.5 million for Bridge Loop structures, 1.5 million for Boolean operations, and 2.4 million for Array-based constructions. In total, our constructed part dataset comprises  around 10 million point cloud–code pairs. We partitioned the dataset into 70\% for training, 15\% for validation, and 15\% for testing. 

\subsubsection{Object Dataset}
We trained our model on the Infinigen Indoor~\cite{infinigen2024indoors} dataset. \textbf{Infinigen Indoor} is a procedural framework for generating synthetic 3D indoor objects, 
where each generated instance is automatically composed by its corresponding parts.
We have made extensive modifications to the original Infinigen codebase to enable it to produce both individual components and their complete assemblies.
Using this framework, we constructed a synthetic dataset comprising 41 common object categories, generating 1 million object-code pairs in total. 
We partitioned the dataset into training, validation, and test sets, following the same split strategy as the Synthetic Part Dataset.  For more details, please refer to the \ref{appendix: datasets}.

\subsection{Implementation Details}
We conduct training and evaluation on the Infinigen Indoor datasets~\cite{Mo_2019_CVPR}. 
For the part-to-code reconstruction model, we adopt the AdamW optimizer and train it for 20 epochs on NVIDIA A100 GPUs with a batch size of 512, and a learning rate of $10^{-4}$.
We evaluate the model at every epoch and select the checkpoint with the lowest $L_2$ Chamfer Distance (CD) loss.
Then we initialize the weights of the object-to-code reconstruction model with the weights of the trained part-to-code reconstruction model,
and train the model on Infinigen Indoor dataset for 10 epochs, with a batch size of 256, and a learning rate of $10^{-4}$.
The checkpoint with the lowest CD loss is selected.
For additional training details and the parameter settings of the models, please refer to ~\ref{appendix: training} and ~\ref{appendix: Structure of the shape tokenizer model}.

\subsection{Reconstruction Performance}
\begin{figure}[ht]
    \vspace{-5pt}
    \centering
    \includegraphics[width=14cm]{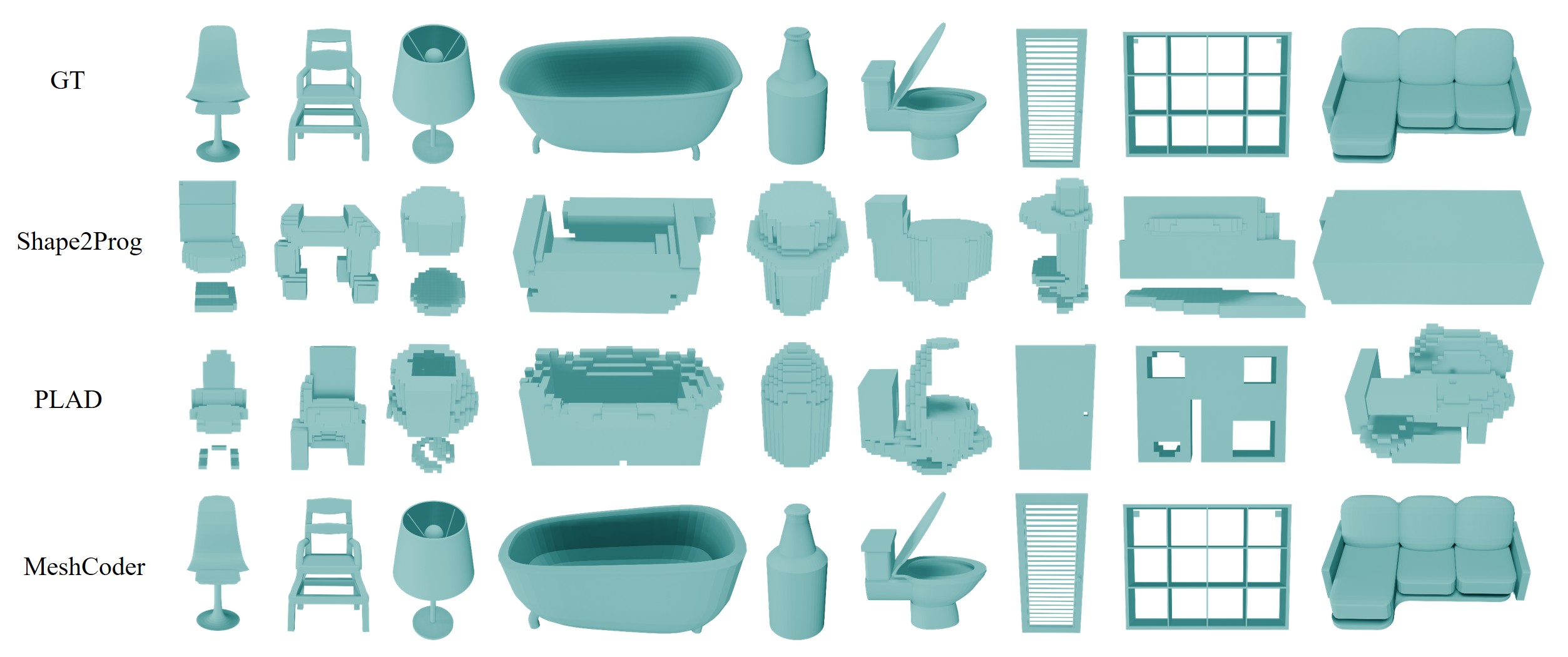}
    \vspace{-10pt}
    \caption{
    Qualitative comparison of reconstruction performance between MeshCoder and baselines.
    MeshCoder can accurately reconstruct objects with intricate parts and complex structures.
    }
    \label{fig:result}
    \vspace{-10pt}
\end{figure}

For reconstruction performance, we compare our method with two representative shape-to-code baselines, Shape2Prog~\cite{tian2018learning} and PLAD~\cite{jones2022plad}. 
Figure~\ref{fig:result} illustrates visualization comparisons of results.
We adopt IoU and $L_2$ CD as our evaluation metrics. Specifically, we voxelize the model's predicted outputs into $32^3$ grids and compute the IoU between the predicted and ground truth voxel grids. In parallel, we sample point clouds from both the predicted outputs and the ground truth, and calculate the Chamfer Distance between the two point clouds.
Regarding the number of points and normalization, please refer to the appendix~\ref{appendix: shape reconstruction}.
In Table ~\ref{tab:infinigen_cd_miou_comparison}, we present reconstruction metrics for some specific object categories as well as the overall performance across the entire dataset. It can be observed that our method consistently outperforms the baselines in both IoU and CD metrics. Complete results for all categories in each dataset are provided in ~\ref{appendix: shape reconstruction}.
We conducted a series of ablation studies to evaluate the impact of various components within our model. For comprehensive details on these experiments,  please refer to ~\ref{appendix: shape reconstruction}.

\begin{table}[ht]
\centering
\caption{
Quantitative comparison of reconstruction performance between MeshCoder and baselines.
}
\label{tab:infinigen_cd_miou_comparison}
\resizebox{\textwidth}{!}{%
\begin{tabular}{l|cccccc|cccccc}
\toprule
\multirow{2}{*}{Method} & \multicolumn{6}{c|}{CD(\( \times 10^{-2} \))↓} & \multicolumn{6}{c}{IoU (\%) ↑} \\
\cmidrule(lr){2-7} \cmidrule(lr){8-13}
 & Lamp & Chair & Sofa & TableDining & Toilet & All & Lamp & Chair & Sofa & TableDining & Toilet & All \\
\midrule
Shape2Prog   & 25.44 & 1.30 & 2.14 & 1.03 & 7.51 & 6.01 & 16.96 & 49.68 & 65.29 & 71.26 & 51.14 & 45.03 \\
PLAD         & 1.40 & 2.26 & 1.52 & 5.52 & 2.30 & 1.87 & 69.58 & 40.93 & 81.33 & 58.43 & 62.61 & 67.62 \\
\textbf{MeshCoder} & \textbf{0.004} & \textbf{0.060} & \textbf{0.027} & \textbf{0.024} & \textbf{0.022} & \textbf{0.063} & \textbf{86.23} & \textbf{81.87} & \textbf{93.81} & \textbf{88.14} & \textbf{89.10} & \textbf{86.75} \\
\bottomrule
\end{tabular}%
}
\end{table}


\subsection{Shape Editing}
\begin{figure}[t]
    \centering
    \includegraphics[width=14cm]{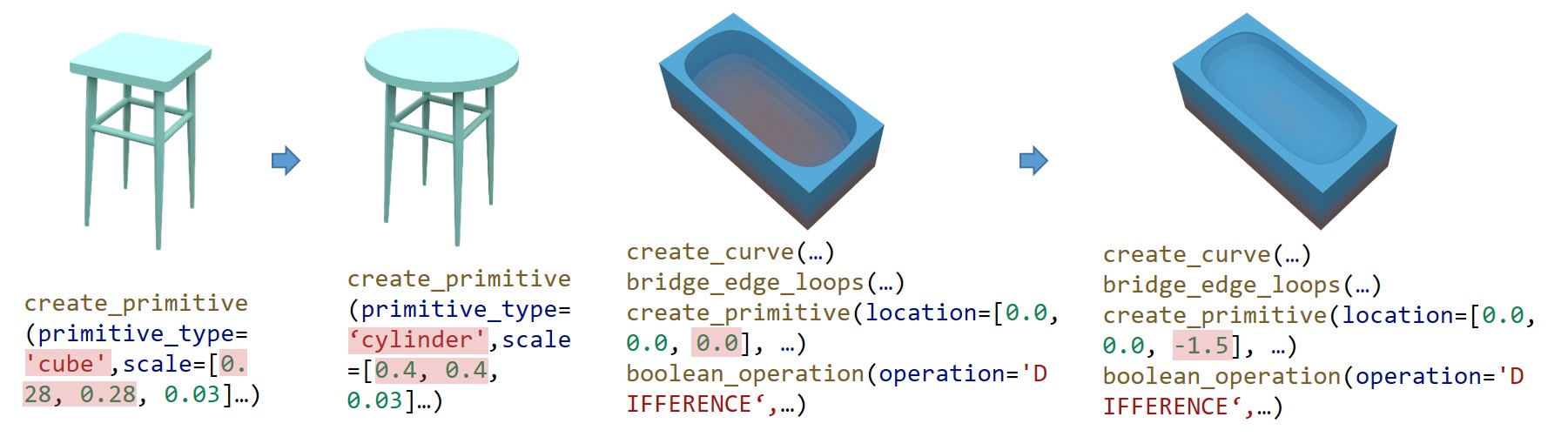}
    \vspace{-10pt}
    \caption{Parameter modification in the code conveniently to alter the geometric shape. Left: Change tabletop from square to circular. Right: Make the bathtub shallower.}
    \label{fig:edit-change}
    \vspace{-10pt}
\end{figure}

\begin{figure}[t]
    \centering
    \includegraphics[width=14cm]{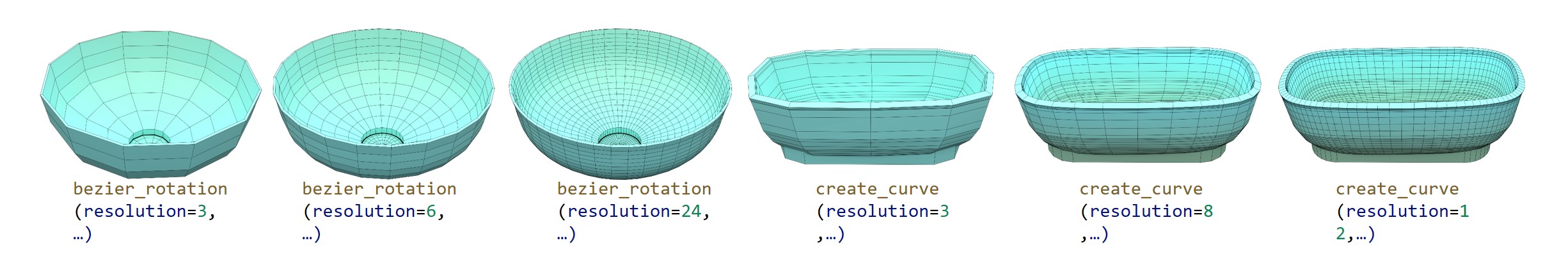}
    \vspace{-10pt}
    \caption{Mesh resolution adjustment by modifying the resolution parameters in the code. 
    The figure depicts results with progressively increasing resolution from left to right.}
    \label{fig:edit}
    \vspace{-10pt}
\end{figure}

MeshCoder facilitates the transformation of 3D shapes into high-level, human-readable code representations, significantly enhancing the interpretability and editability of complex geometries. This capability enables intuitive and precise modifications through code-based interventions. Our shape editing encompasses two primary categories: geometric editing and topological editing. As illustrated in Figure~\ref{fig:edit-change}, geometric editing can be performed by adjusting function calls or modifying specific parameters within the generated code. For instance, we can adjust the parameters of the code to convert a square tabletop into a larger circular one. Additionally, topological editing, which is illustrated in Figure~\ref{fig:edit} such as adjusting mesh resolution, can be achieved by modifying designated parameters within the code, allowing for control over the mesh's complexity and surface detail. This code-centric approach streamlines the process of modifying 3D models, making it more accessible and efficient for applications requiring iterative design and customization. Additionally, it empowers users to adjust the model resolution according to their desired balance between storage requirements and mesh quality. For additional results and details, please refer to \ref{appendix: Shape Editing} in the appendix.

\subsection{Shape Undertanding}

\begin{wrapfigure}{r}{0.5\textwidth}
    \vspace{-0.3cm}
    \centering
    \includegraphics[width=0.5\textwidth]{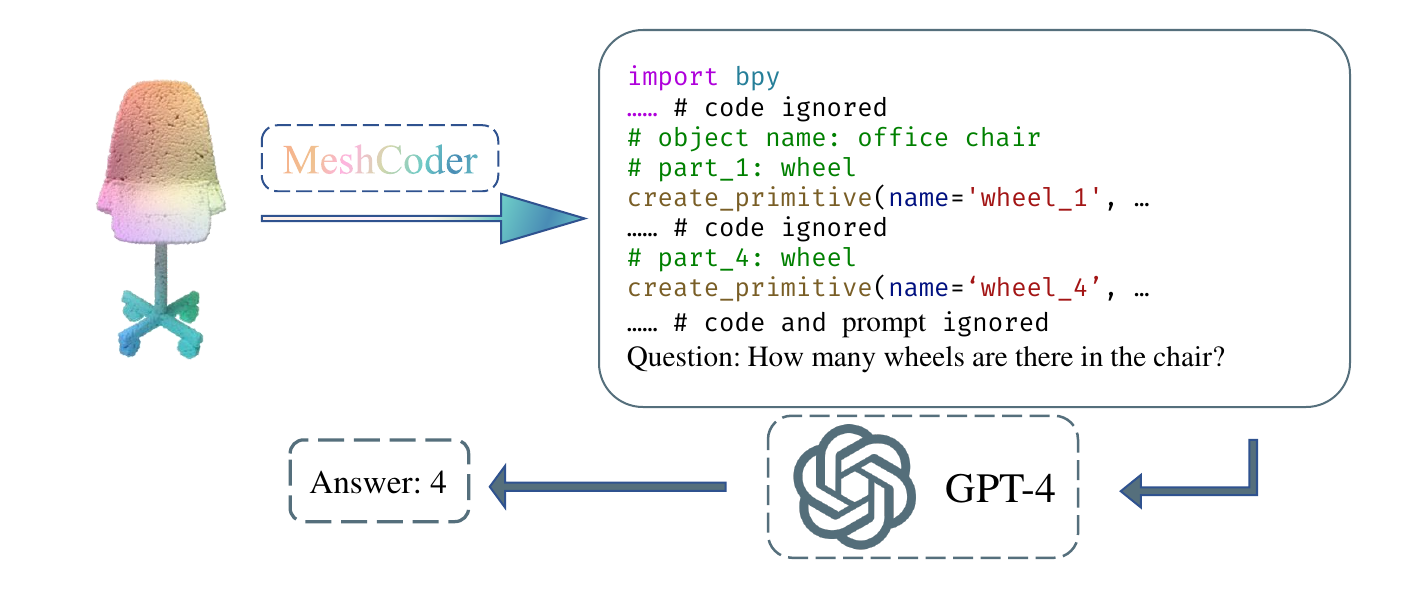}
    \captionsetup{width=0.45\textwidth}
    \caption{The pipeline of conducting experiments on shape understanding. }
    \label{fig:understanding}
    \vspace{-0.4cm}
\end{wrapfigure}

MeshCoder is capable of predicting object codes with rich semantic information. These codes effectively capture structural and geometric details, making them valuable for shape understanding. By inputting the predicted codes into GPT, we can assist it in comprehending object structures. We conduct experiments on shape understanding, with an example illustrated in Figure~\ref{fig:understanding}.
Additional results and details are given in \ref{appendix: Shape Undertanding} in the appendix.

\section{Limitations}
\label{sec:limitation} 
Although our method achieves significant advancements in category diversity, geometric complexity, and reconstruction accuracy compared to existing approaches, it primarily targets human-made objects. The applicability of code-based representations to organic forms, such as animals and humans, remains underdeveloped. We reserve this as a direction for future research.

\section{Conclusion}
\label{sec:conclusion}
In this work, we present MeshCoder, a comprehensive framework that translates 3D point cloud data into editable Blender Python scripts, enabling detailed reconstruction and intuitive editing of complex 3D objects. By developing a robust set of Blender Python APIs, we facilitate the modeling of intricate geometries. Leveraging these APIs, we constructed a large-scale dataset pairing 3D objects with their corresponding code representations, decomposed into semantic parts. Subsequently, we trained a multimodal large language model (LLM) capable of generating executable Blender scripts from point cloud inputs.
Our approach not only achieves superior performance in shape-to-code reconstruction tasks but also enhances the reasoning capabilities of LLMs in 3D shape understanding. By representing shapes as structured code, MeshCoder offers a flexible and powerful solution for programmatic 3D shape reconstruction and editing, paving the way for advanced applications in reverse engineering, design, and analysis.


\bibliographystyle{unsrtnat}
\bibliography{main}
\newpage
\appendix

\section{Appendix of MeshCoder: LLM-Powered Structured Mesh Code Generation from Point Clouds}
\subsection{Datasets}
\label{appendix: datasets}

\subsubsection{The principles of Translation and Bridge Loop}
\begin{figure}[h]
    \centering
    \includegraphics[width=14cm]{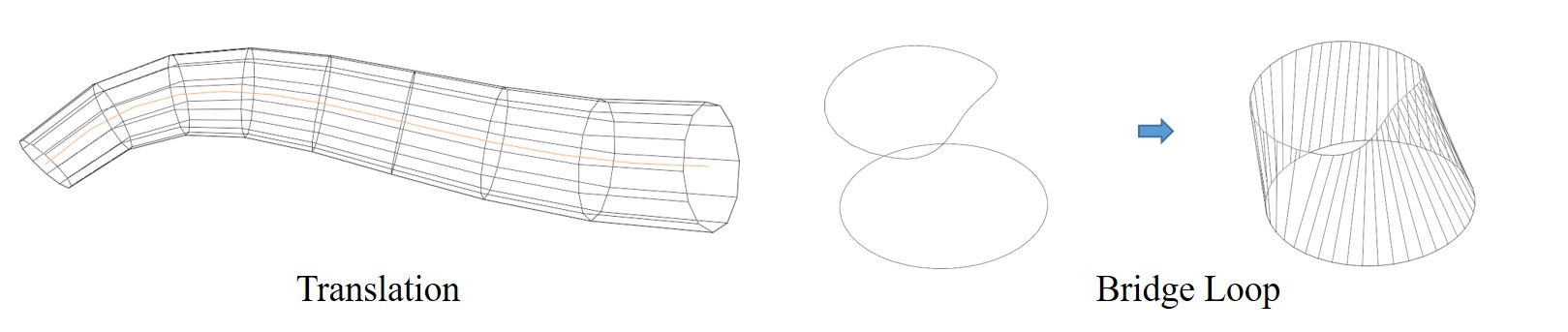}
    \caption{
    A schematic illustration of the principles of Translation and Bridge Loop. In the Translation module, the wireframe of the resulting mesh is shown as a cross-sectional circle is translated along a yellow trajectory. In the Bridge Loop module, the wireframe of the mesh is constructed by connecting the vertices of two 2D shapes.}
    \label{fig:trans_and_bridge}
\end{figure}

As illustrated in the figure~\ref{fig:trans_and_bridge}, in the Translation operation, a 2D cross-sectional shape (a circle in this example) and a 3D trajectory curve must first be defined. The Translation process generates a mesh by sweeping the 2D shape along the 3D trajectory. During this sweep, the cross-section remains perpendicular to the tangent direction of the trajectory at all times, and only uniform scaling (either enlargement or reduction) of the cross-section is permitted.

In contrast, the Bridge Loop operation begins with two predefined 2D shapes. By connecting the corresponding vertices of these two shapes, a mesh can be constructed. This method places no constraints on the types of 2D shapes used—meaning the two shapes can differ, such as a circle and a irregular closed shape in this example. Moreover, it imposes no restrictions on the relative orientations of the shapes. As a result, Bridge Loop overcomes the limitations of Translation, which requires the cross-section to align with the trajectory’s tangent direction. This enables Bridge Loop to generate more complex geometries that Translation cannot produce.

\subsubsection{Part datasets}

\begin{figure}[ht]
    \centering
    \includegraphics[width=14cm]{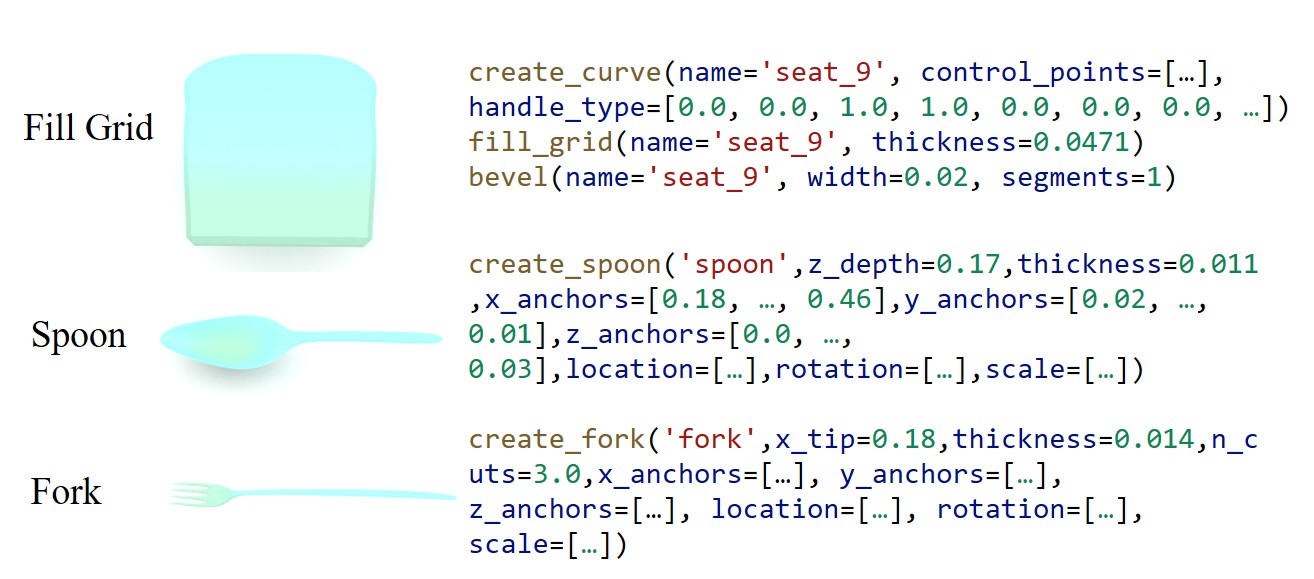}
    \caption{
    The Fill Grid type, Spoon type and Fork type in basic shape code library}
    \label{fig:unique}
\end{figure}

For certain shapes that are difficult to represent using the method we defined in Section~\ref{sec: part_and_code_dataset}, we introduce three additional categories: the Fill Grid type, Spoon type and Fork type. As illustrated in the Figure~\ref{fig:unique}. For the Fill Grid type, we first construct a closed 3D shape (as opposed to the 2D cross-sectional shape used in Translation), fill it to form a surface, and then extrude it along its normal direction to generate the final mesh. For the Spoon and Fork type, we draw inspiration from the implementation in Infinigen Indoor~\cite{infinigen2024indoors} and design dedicated procedural functions tailored for their generation.

\begin{figure}[ht]
    \centering
    \includegraphics[width=14cm]{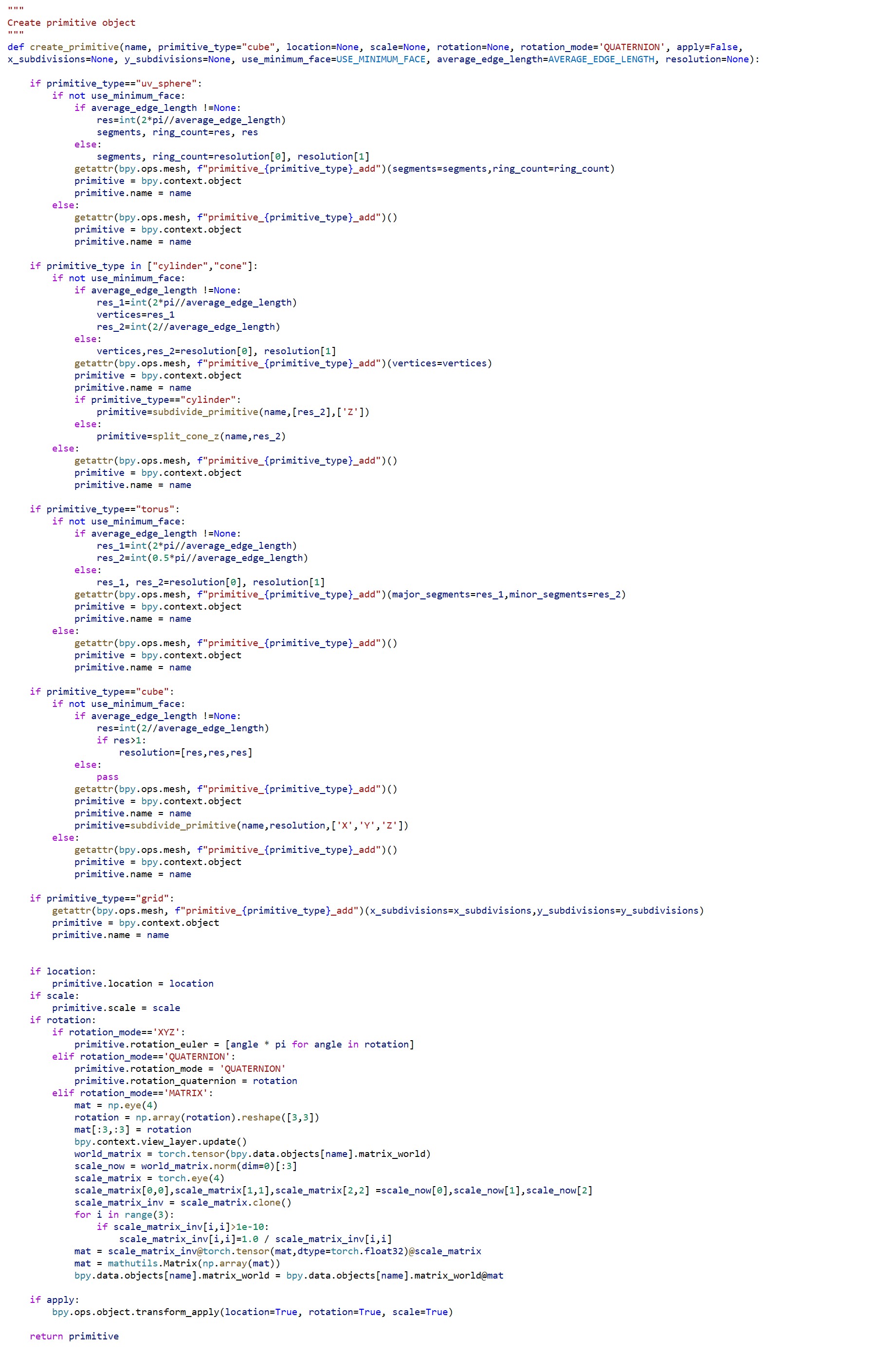}
    \caption{
    Implementation of the function for creating primitives}
    \label{fig:code_primitive}
\end{figure}

We present two core functions from our codebase: the complete implementation for creating primitives (Figure~\ref{fig:code_primitive}) and the complete implementation for creating curves (Figure~\ref{fig:code_curve}). The full codebase can be found in the supplementary materials.

\begin{figure}[ht]
    \centering
    \includegraphics[width=14cm]{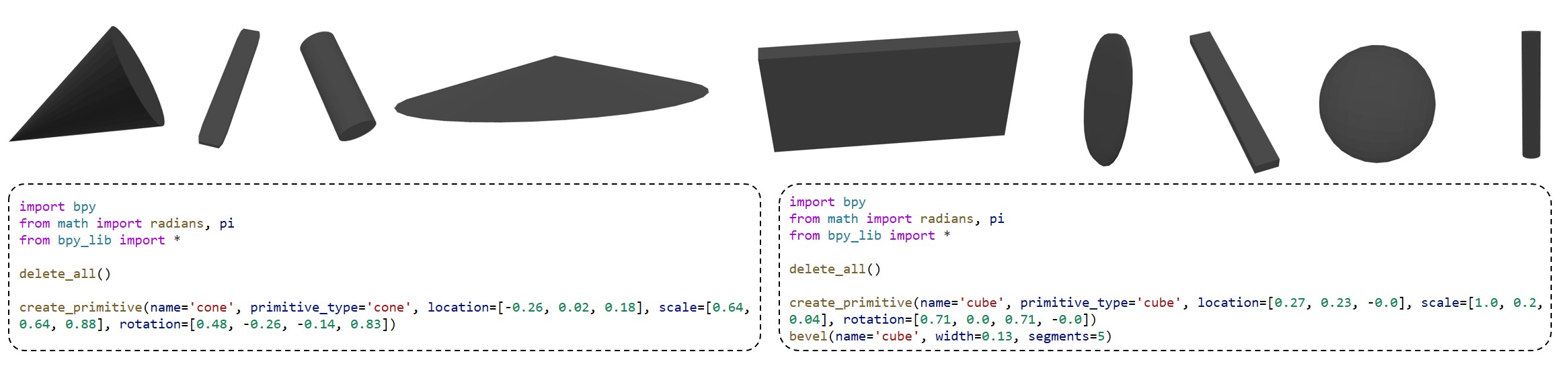}
    \caption{
    Examples of Primitive and complete code. And the code corresponds to the first two objects shown in the figure.}
    \label{fig:primitive_example}
\end{figure}

\begin{figure}[ht]
    \centering
    \includegraphics[width=14cm]{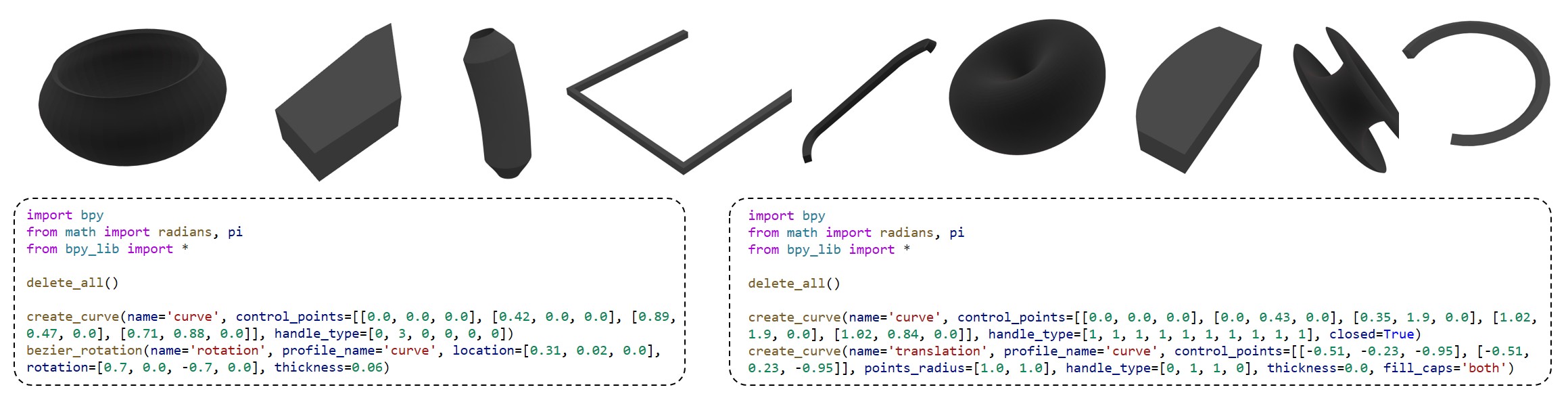}
    \caption{
    Examples of Translation and complete code. And the code corresponds to the first two objects shown in the figure.} 
    \label{fig:translation_example}
\end{figure}

\begin{figure}[ht]
    \centering
    \includegraphics[width=14cm]{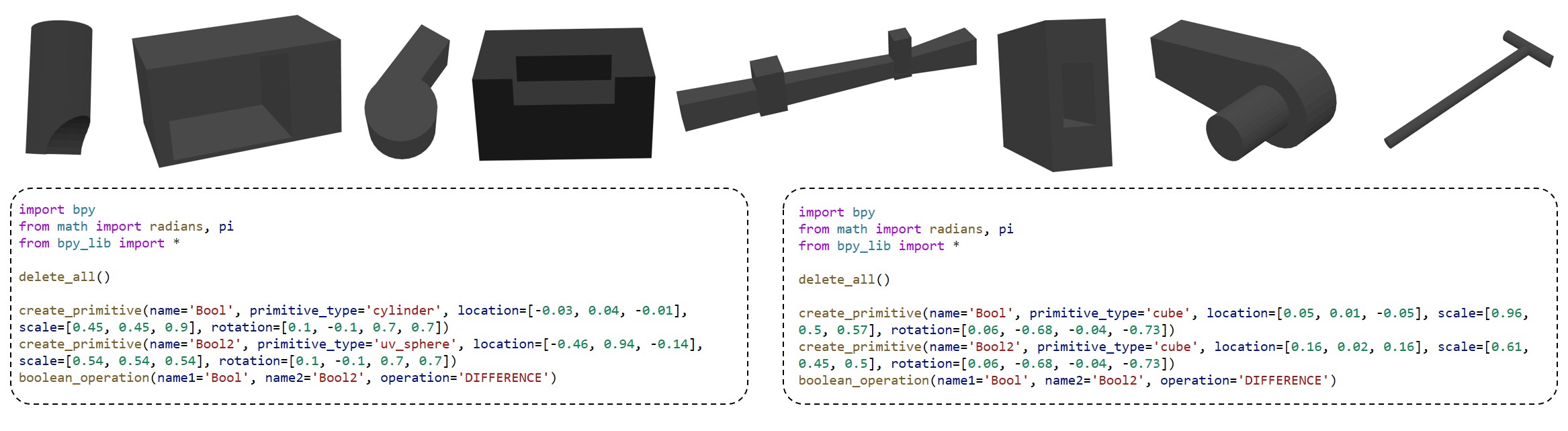}
    \caption{
    Examples of Boolean and complete code. And the code corresponds to the first two objects shown in the figure.}
    \label{fig:bool_example}
\end{figure}

\begin{figure}[ht]
    \centering
    \includegraphics[width=14cm]{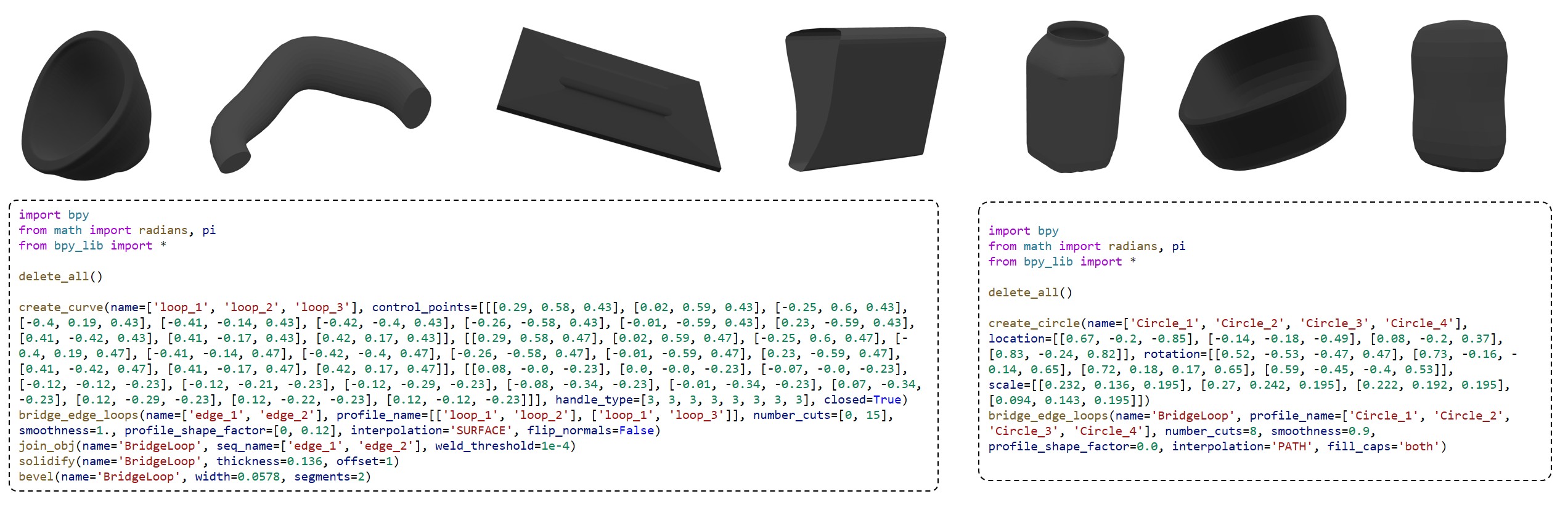}
    \caption{
    Examples of Bridge Loop and complete code. And the code corresponds to the first two objects shown in the figure.}
    \label{fig:bridge_example}
\end{figure}

\begin{figure}[ht]
    \centering
    \includegraphics[width=14cm]{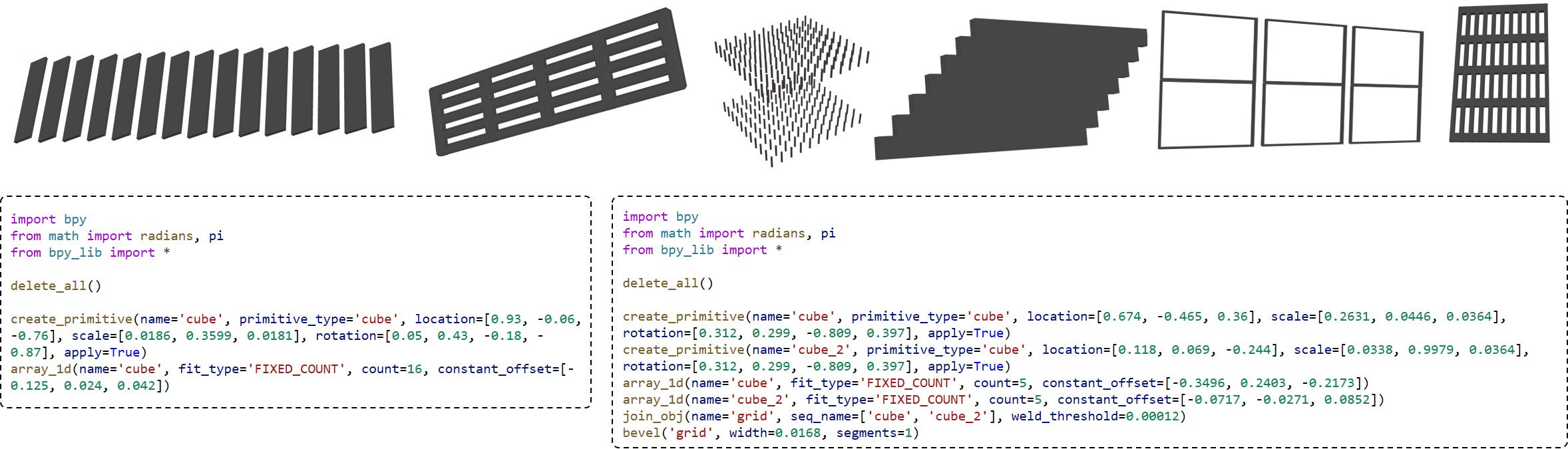}
    \caption{
    Examples of Array and complete code. And the code corresponds to the first two objects shown in the figure.}
    \label{fig:array_example}
\end{figure}

More examples of parts and their corresponding complete code implementations are provided in Figures~\ref{fig:primitive_example},~\ref{fig:translation_example},~\ref{fig:bool_example},~\ref{fig:bridge_example}, and~\ref{fig:array_example}.

Taking the \texttt{Primitive} type as an example, we describe how to use functions from the basic shape code library to generate a synthetic part dataset. We begin by randomly selecting the type of primitive to generate (e.g., \texttt{cube}, \texttt{cylinder}, etc.). Next, for each axis, we independently uniform sample a value \( x \) from the range \([-2, 2]\), and then set the corresponding scale as \( 10^{x} \).
To determine the orientation of the shape, we uniformly sample a direction from a unit sphere and a roll angle from a uniform distribution. Once the orientation is fixed, we scale the shape uniformly along all three axes based on the size of its bounding box. Specifically, we ensure that the longest edge of the bounding box lies within the range $[1, 2]$.
Finally, we assign the shape a random position within the 3D space such that the entire shape remains within the $[-1, 1]$ bounds. For other shape types beyond \texttt{Primitive}, we follow a similar approach by randomly assigning values to the relevant parameters.

\subsubsection{Object datasets}
\begin{figure}[ht]
    \centering
    \includegraphics[width=14cm]{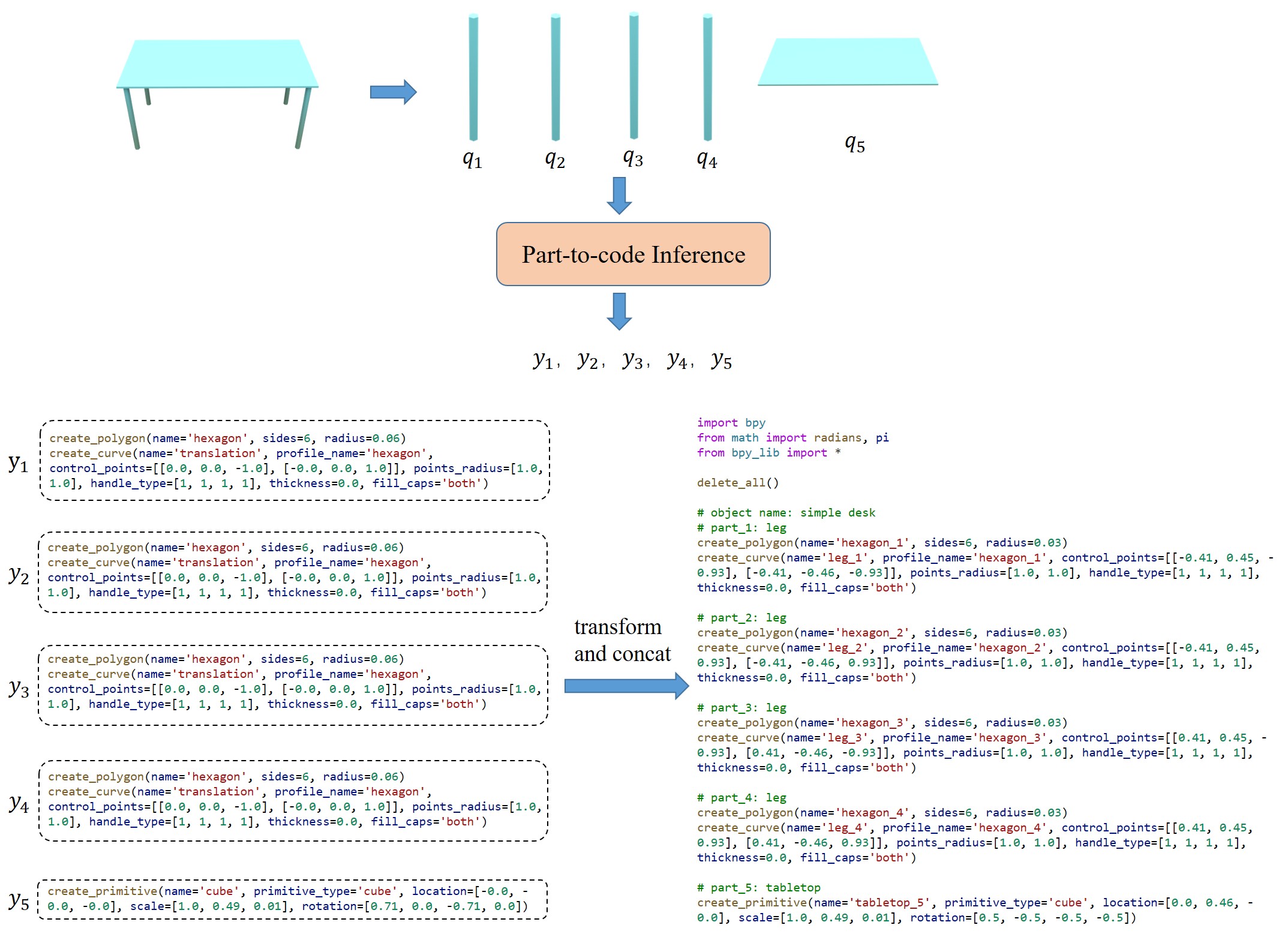}
    \caption{
    A complete code example of converting part codes into a full object program.
    }
    \label{fig:full_code_concat.}
\end{figure}

For assembling part codes into a complete program, we provide a full example containing the complete code, as shown in Figure~\ref{fig:full_code_concat.}. Regarding the ordering strategy used when assembling parts into a complete object, we adopt a consistent spatial heuristic to determine part sequence. Specifically, parts are arranged from bottom to top, left to right, and front to back. To implement this, we divide the 3D space into a $32 \times 32 \times 32$ grid and assign each part a characteristic grid cell that serves as the basis for sorting.
The characteristic grid cell of a part is defined as follows: among all grid cells that the part occupies, we first select the one with the smallest $z$-coordinate. If multiple candidates share the same $z$-value, we choose the one with the smallest $x$-coordinate. If a tie still exists, we select the one with the smallest $y$-coordinate. Parts are then sorted based on the lexicographic order of these characteristic grid cells, which determines their final sequence within the object.

It is important to note that for each object, the prerequisite for successfully constructing its corresponding code lies in the ability of our part-to-code inference model to accurately infer all of its individual parts. We consider a part to be successfully inferred if the Chamfer Distance (CD) between the predicted point cloud and the ground truth is below $5 \times 10^{-3}$. Therefore, when constructing the object-code pairs dataset, we only include objects for which \textbf{all} constituent parts meet this criterion. Objects with any part failing to meet this standard are discarded. As a result, the number of successfully constructed object-code pairs is smaller than the total number of objects in the original Infinigen dataset. 
In fact, the original Infinigen dataset we use contains 1.57 million object instances, from which we successfully construct 1 million shape-code pairs. For training and evaluation, we split the full Infinigen dataset into 70\% for training, 15\% for testing, and 15\% for validation. 
Accordingly, \textsc{MeshCoder} is trained only on the subset of the shape-code pairs that fall within the training portion of the Infinigen dataset. In contrast, the baseline models are trained on the full set of objects in the training split of the original Infinigen dataset.
Importantly, all evaluation results for our method and the baselines are reported on the same test set, i.e., the testing split of the complete Infinigen dataset.

\subsection{Model architecture}
\label{appendix: Structure of the shape tokenizer model}
\begin{figure}[ht]
    \centering
    \includegraphics[width=14cm]{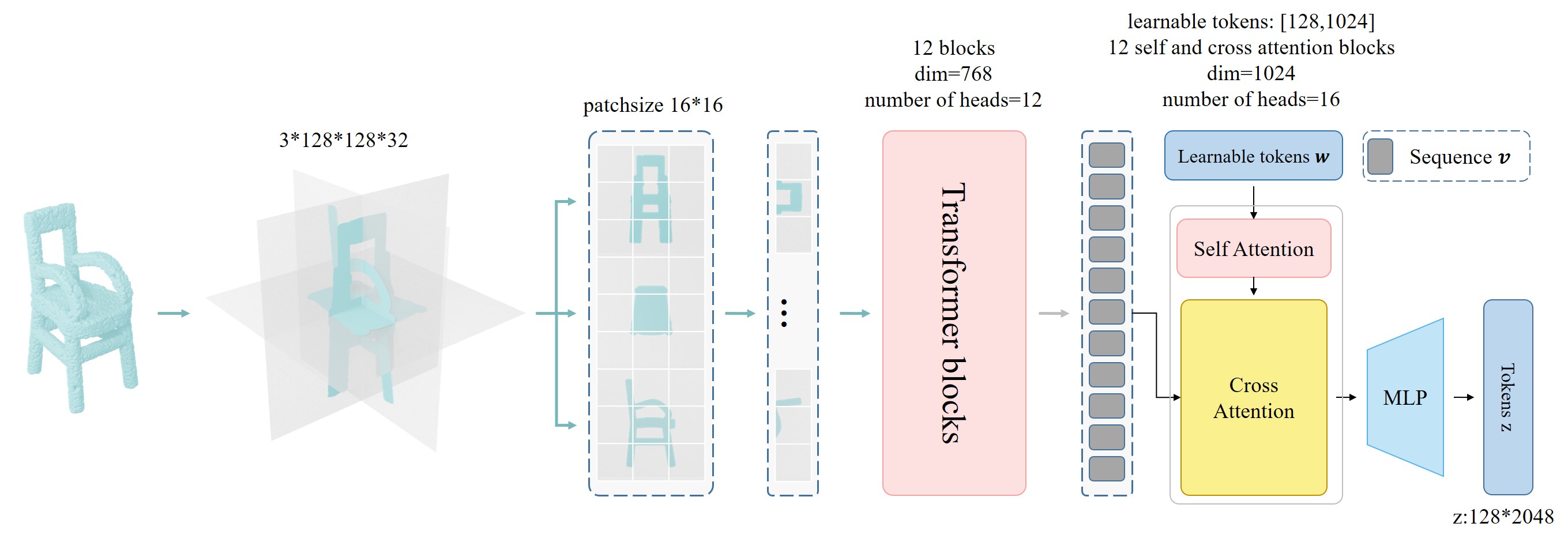}
    \caption{
    Detailed configuration of the shape tokenizer.}
    \label{fig:configuration of the shape tokenizer.}
\end{figure}
We explain the detailed structure of the shape tokenizer.
As illustrated in the Figure~\ref{fig:configuration of the shape tokenizer.}, we first project the input point cloud of shape $\mathbb{R}^{n \times 3}$ onto three orthogonal planes to obtain tri-plane features with shape $\mathbb{R}^{3 \times 128 \times 128 \times 32}$. With a patch size set to $16 \times 16$, these tri-plane features are encoded into tokens and fed into Transformer blocks, where the resulting representation is mapped to $v$ and used as the key and value ($\mK$, $\mV$) inputs.
Meanwhile, a set of learnable tokens with shape $\mathbb{R}^{128 \times 1024}$ are used as queries in a self and cross attention module. After passing through 12 layers of self and cross attention, we obtain output tokens of shape $\mathbb{R}^{128 \times 1024}$, which are then projected to the final representation of shape $\mathbb{R}^{128 \times 2048}$ via an MLP.

\subsection{More training details}
\label{appendix: training}

For the part-to-code reconstruction model, we adopt the AdamW optimizer and train it for 20 epochs on 64 NVIDIA A100 GPUs for about a week with a batch size of 512, and a learning rate of $10^{-4}$.
We evaluate the model at every epoch and select the checkpoint with the lowest $L_2$ Chamfer Distance (CD) loss.
Then we initialize the weights of the object-to-code reconstruction model with the weights of the trained part-to-code reconstruction model,
and train the model on Infinigen Indoor dataset for 10 epochs, with a batch size of 256, and a learning rate of $10^{-4}$.
It is trained on 64 NVIDIA A100 GPUs for about 2 days.
The checkpoint with the lowest CD loss is selected.

To further enhance the robustness and generalization ability of the object-to-code inference model, we apply data augmentation techniques. Specifically, we perform random rotation and scaling on the objects. Additionally, during training, we randomly sample the number of points in each point cloud within the range of 4096 to 16384, and add Gaussian noise to further perturb the input. MeshCoder is trained and evaluated on a unified dataset that aggregates all object categories.

\subsection{Complete experiment result of Shape Reconstruction}
\label{appendix: shape reconstruction}

For \textbf{MeshCoder}, during inference, each object is represented by a point cloud containing 16{,}384 points. Given the input point cloud, the object-to-code inference model is able to predict the corresponding Blender Python script code. The resulting code is then executed to generate a corresponding mesh. We uniformly sample 100{,}000 points from the generated mesh and compute the Chamfer Distance (CD) to the input point cloud using the $L_2$ norm.

Given two point sets $P$ and $Q$, each of size 100{,}000, the $L_2$ Chamfer Distance is defined as:
\[
\text{CD}(P, Q) = \frac{1}{|P|} \sum_{x \in P} \min_{y \in Q} \|x - y\|_2^2 + \frac{1}{|Q|} \sum_{y \in Q} \min_{x \in P} \|y - x\|_2^2.
\]

To evaluate IoU, we voxelize both the ground-truth mesh and the predicted mesh into grids of resolution $32^3$, and compute the voxel-based Intersection-over-Union (IoU) as:
\[
\text{IoU} = \frac{|\mathcal{V}_\text{pred} \cap \mathcal{V}_\text{gt}|}{|\mathcal{V}_\text{pred} \cup \mathcal{V}_\text{gt}|},
\]
where $\mathcal{V}_\text{pred}$ and $\mathcal{V}_\text{gt}$ denote the sets of occupied voxels in the predicted and ground-truth voxel grids, respectively.

For \textbf{baseline methods}, which take voxel grids as input and output voxel grids, we first voxelize the ground-truth mesh into a $32^3$ grid and feed it into the baseline models. The predicted voxel grid is then compared to the input voxelized ground truth to compute IoU. Additionally, we extract a mesh from the predicted voxel grid using the Marching Cubes algorithm and uniformly sample 100{,}000 points from the resulting mesh surface. These sampled points, along with the ground-truth point cloud, are then both uniformly scaled to fit within the $[-1, 1]^3$ volume. Finally, the Chamfer Distance is computed between the two normalized point clouds using the $L_2$ norm.

It's noticed that for each object category, we independently train the baseline models, according to their official code, resulting in category-specific checkpoints. These models are then evaluated on the corresponding test sets for each category. 

\begin{figure}[ht]
    \centering
    \includegraphics[width=14cm]{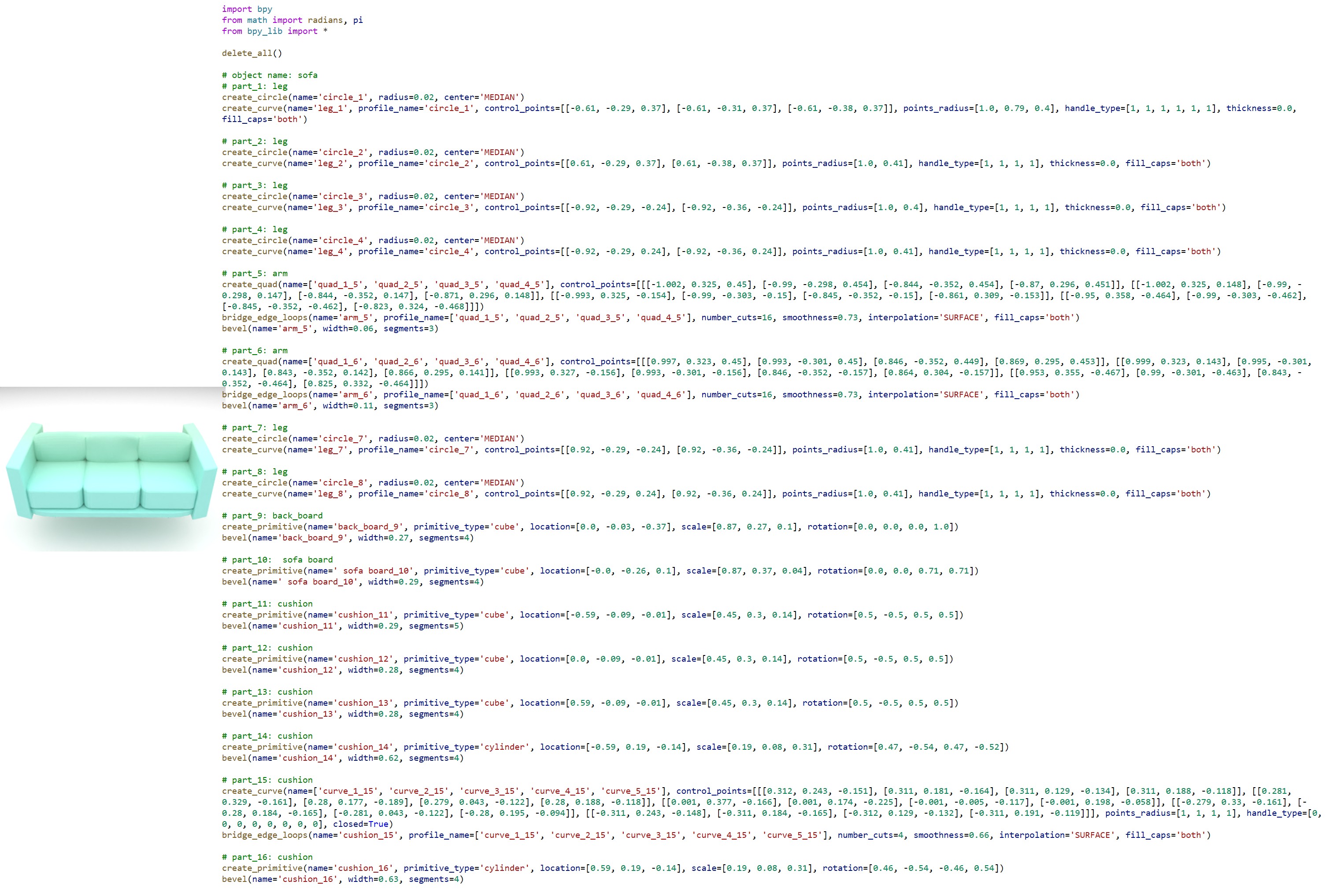}
    \caption{
    An example of sofa. The input is a point cloud of a sofa, and the figure shows the code inferred by the object-to-code inference model, as well as the resulting mesh generated by executing the inferred code.}
    \label{fig:sofa_example.}
\end{figure}
\begin{figure}[ht]
    \centering
    \includegraphics[width=14cm]{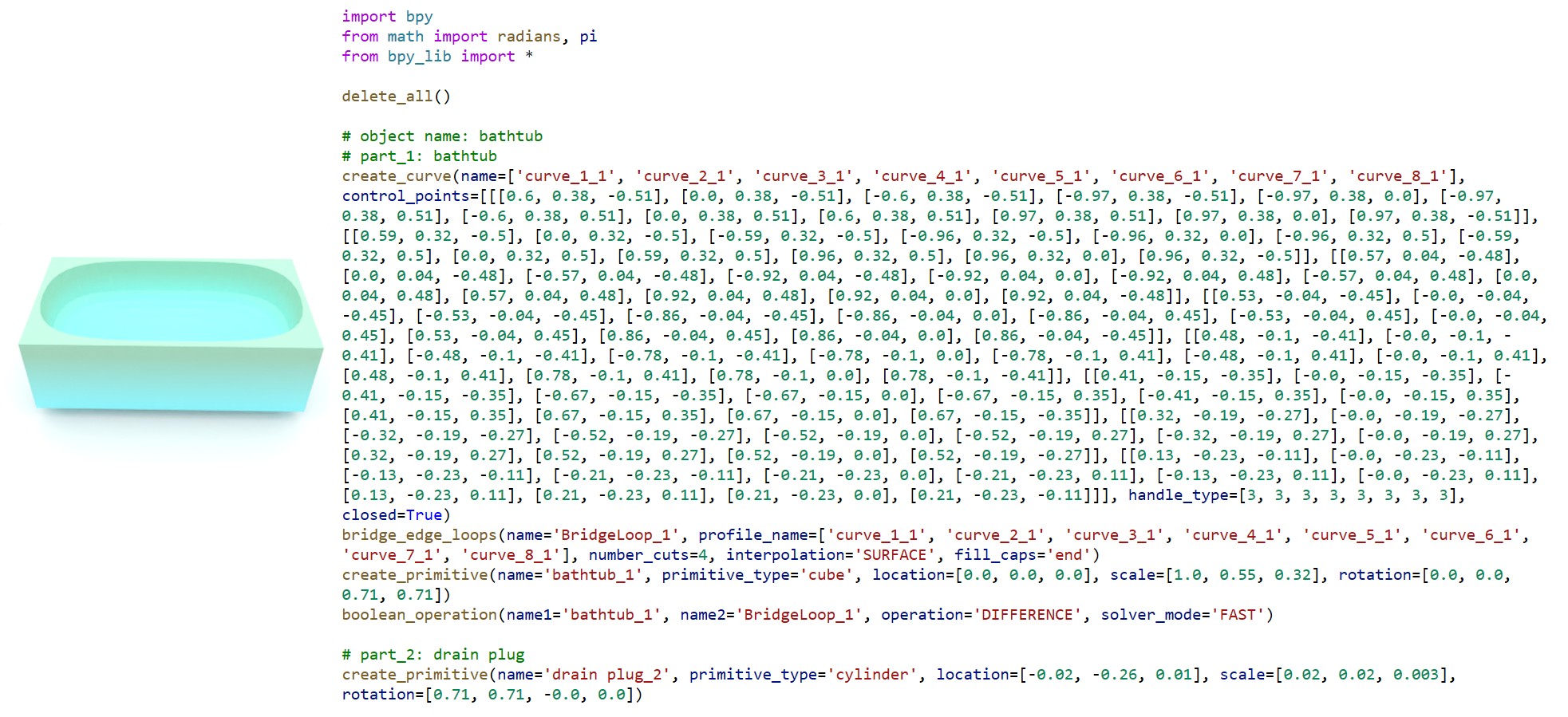}
    \caption{
    An example of bathtub. The input is a point cloud of a bathtub, and the figure shows the code inferred by the object-to-code inference model, as well as the resulting mesh generated by executing the inferred code.}
    \label{fig:bathtub_example.}
\end{figure}
\begin{figure}[ht]
    \centering
    \includegraphics[width=14cm]{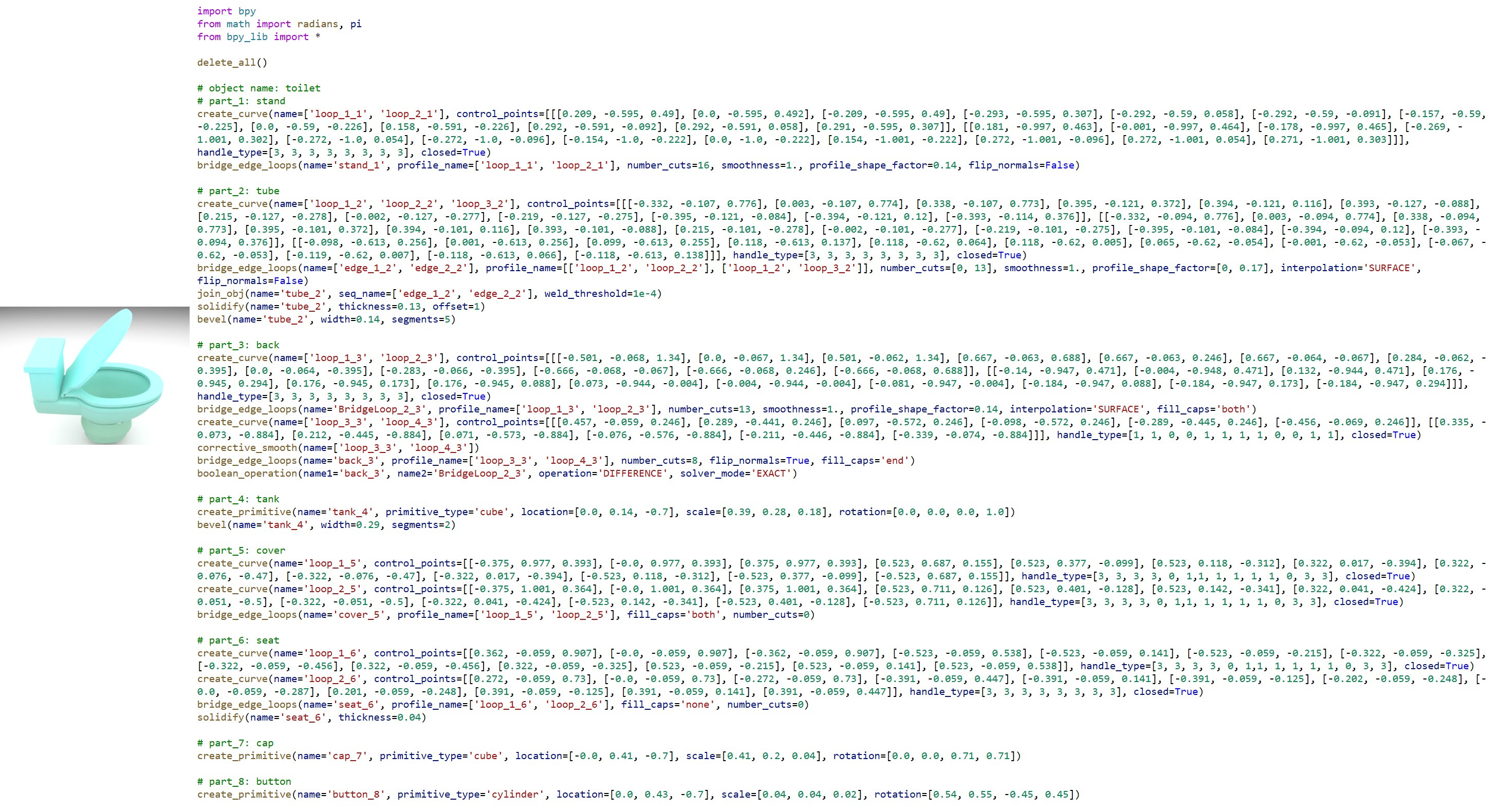}
    \caption{
    An example of toilet. The input is a point cloud of a toilet, and the figure shows the code inferred by the object-to-code inference model, as well as the resulting mesh generated by executing the inferred code.}
    \label{fig:toilet_example.}
\end{figure}

The quantitative comparison of reconstruction metrics between MeshCoder and baseline methods across all object categories is summarized in Table~\ref{tab:reconstruction_metrics} and Table~\ref{tab:reconstruction_std_metrics}.
Some additional examples of object reconstruction results and their complete code can be referred to Figure \ref{fig:sofa_example.}, \ref{fig:bathtub_example.}, \ref{fig:toilet_example.}.

\begin{table*}[t]
\centering
\caption{Comparison of reconstruction metrics across all categories. Chamfer Distance (CD) and IoU is shown in percentage (\%).}
\label{tab:reconstruction_metrics}
\resizebox{\textwidth}{!}{
\begin{tabular}{l|ccc|ccc}
\toprule
\textbf{Category} & \multicolumn{3}{c|}{\textbf{L2 CD($\times 10^{-2}$})} & \multicolumn{3}{c}{\textbf{Voxel IoU (\%)}} \\
 & MeshCoder & PLAD & Shape2prog & MeshCoder & PLAD & Shape2prog \\
\midrule
ArmChair & 0.04 & 2.31 & 4.44 & 94.33 & 78.79 & 62.74 \\
BarChair & 0.03 & 2.23 & 2.55 & 88.73 & 74.96 & 58.23 \\
Bathtub & 0.09 & 1.22 & 2.45 & 78.70 & 74.50 & 42.94 \\
BeverageFridge & 0.22 & 1.12 & 12.63 & 88.03 & 82.13 & 39.13 \\
Bottle & 0.01 & 1.08 & 6.34 & 88.65 & 65.58 & 40.24 \\
Bowl & 0.02 & 1.43 & 6.29 & 89.93 & 60.02 & 25.60 \\
CeilingClassicLamp & 0.02 & 1.98 & 3.94 & 96.13 & 76.01 & 59.07 \\
CeilingLight & 0.03 & 3.46 & 1.32 & 65.83 & 40.61 & 44.97 \\
CellShelf & 0.01 & 1.93 & 9.40 & 94.67 & 59.02 & 22.30 \\
Chair & 0.06 & 2.26 & 1.30 & 81.87 & 40.93 & 49.68 \\
Chopsticks & 0.03 & 1.38 & 21.06 & 82.24 & 55.68 & 11.25 \\
Cup & 0.06 & 1.40 & 7.35 & 85.96 & 62.03 & 29.47 \\
DeskLamp & 0.02 & 1.76 & 8.77 & 80.28 & 64.31 & 25.35 \\
Dishwasher & 0.13 & 1.44 & 3.01 & 88.37 & 84.44 & 46.69 \\
FloorLamp & 0.00 & 2.13 & 22.97 & 85.96 & 66.89 & 17.16 \\
Fork & 0.14 & 0.34 & 8.40 & 58.86 & 89.28 & 11.03 \\
Hardware & 0.01 & 0.62 & 8.45 & 89.87 & 83.96 & 23.56 \\
Jar & 0.03 & 0.76 & 1.39 & 79.12 & 69.67 & 41.51 \\
Lamp & 0.00 & 1.40 & 25.44 & 86.23 & 69.58 & 16.96 \\
LargeShelf & 0.02 & 0.82 & 5.15 & 88.08 & 60.70 & 16.81 \\
Lid & 0.05 & 1.83 & 2.39 & 73.22 & 63.47 & 50.11 \\
LiteDoor & 0.03 & 1.36 & 5.75 & 94.75 & 36.91 & 18.71 \\
LouverDoor & 0.07 & 1.40 & 16.17 & 89.46 & 37.43 & 20.94 \\
Microwave & 0.07 & 1.44 & 11.04 & 91.72 & 55.65 & 49.38 \\
OfficeChair & 0.03 & 1.44 & 2.63 & 78.41 & 55.65 & 46.91 \\
PanelDoor & 0.04 & 1.31 & 6.50 & 94.60 & 37.18 & 20.94 \\
Plate & 0.04 & 0.96 & 1.07 & 72.70 & 70.72 & 60.05 \\
SidetableDesk & 0.01 & 0.67 & 4.50 & 93.23 & 91.75 & 35.75 \\
SimpleBookcase & 0.03 & 1.78 & 2.89 & 92.14 & 65.14 & 33.79 \\
SimpleDesk & 0.01 & 2.12 & 25.39 & 88.68 & 93.80 & 45.79 \\
Sofa & 0.03 & 1.52 & 2.14 & 93.81 & 81.33 & 65.29 \\
Spoon & 0.67 & 0.37 & 4.09 & 74.00 & 87.04 & 18.92 \\
TableCocktail & 0.02 & 2.59 & 5.93 & 88.47 & 60.49 & 25.19 \\
TableDining & 0.02 & 5.52 & 1.03 & 88.14 & 58.43 & 71.26 \\
Toilet & 0.02 & 2.30 & 7.51 & 89.10 & 62.61 & 51.14 \\
TriangleShelf & 0.01 & 2.30 & 12.61 & 88.75 & 62.61 & 30.59 \\
TV & 0.04 & 1.53 & 3.41 & 87.80 & 72.69 & 34.14 \\
TVStand & 0.01 & 0.78 & 13.50 & 91.26 & 73.78 & 22.57 \\
Vase & 0.30 & 0.73 & 19.10 & 72.26 & 89.95 & 60.94 \\ 
Window & 0.14 & 0.59 & 3.73 & 87.36 & 84.21 & 64.64 \\
Wineglass & 0.06 & 0.98 & 6.83 & 88.36 & 73.96 & 28.56 \\
\midrule
\textbf{All (Avg.)} & \textbf{0.06} & \textbf{1.87} & \textbf{6.00} & \textbf{86.75} & \textbf{67.62} & \textbf{45.03} \\
\bottomrule
\end{tabular}
}
\end{table*}

\begin{table*}[t]
\centering
\caption{Comparison of standard deviation of reconstruction metrics across all categories.}
\label{tab:reconstruction_std_metrics}
\resizebox{\textwidth}{!}{
\begin{tabular}{l|ccc|ccc}
\toprule
\textbf{Category} & \multicolumn{3}{c|}{\textbf{CD}} & \multicolumn{3}{c}{\textbf{IoU}} \\
& \textbf{MeshCoder} & \textbf{PLAD} & \textbf{Shape2prog} & \textbf{MeshCoder} & \textbf{PLAD} & \textbf{Shape2prog} \\
\midrule
ArmChair & $1.51 \times 10^{-3}$ & $9.35 \times 10^{-3}$ & $1.51 \times 10^{-2}$ & $4.62 \times 10^{-2}$ & $6.48 \times 10^{-2}$ & $5.28 \times 10^{-2}$ \\
BarChair & $1.82 \times 10^{-4}$ & $1.18 \times 10^{-2}$ & $9.90 \times 10^{-3}$ & $8.19 \times 10^{-2}$ & $1.00 \times 10^{-1}$ & $8.30 \times 10^{-2}$ \\
Bathtub & $6.93 \times 10^{-4}$ & $1.02 \times 10^{-2}$ & $6.70 \times 10^{-3}$ & $1.31 \times 10^{-1}$ & $1.91 \times 10^{-1}$ & $8.57 \times 10^{-2}$ \\
BeverageFridge & $4.84 \times 10^{-3}$ & $2.81 \times 10^{-3}$ & $3.44 \times 10^{-2}$ & $1.13 \times 10^{-1}$ & $6.23 \times 10^{-2}$ & $6.64 \times 10^{-2}$ \\
Bottle & $7.56 \times 10^{-5}$ & $6.80 \times 10^{-3}$ & $6.00 \times 10^{-2}$ & $1.13 \times 10^{-1}$ & $7.05 \times 10^{-2}$ & $6.90 \times 10^{-2}$ \\
Bowl & $4.83 \times 10^{-5}$ & $5.13 \times 10^{-3}$ & $8.78 \times 10^{-3}$ & $8.11 \times 10^{-2}$ & $6.24 \times 10^{-2}$ & $2.35 \times 10^{-2}$ \\
CeilingClassicLamp & $7.33 \times 10^{-7}$ & $7.66 \times 10^{-4}$ & $9.86 \times 10^{-4}$ & $3.39 \times 10^{-5}$ & $2.96 \times 10^{-3}$ & $3.82 \times 10^{-2}$ \\
CeilingLight & $1.79 \times 10^{-6}$ & $3.90 \times 10^{-3}$ & $4.44 \times 10^{-3}$ & $3.10 \times 10^{-2}$ & $5.08 \times 10^{-2}$ & $6.93 \times 10^{-2}$ \\
CellShelf & $3.37 \times 10^{-5}$ & $1.94 \times 10^{-2}$ & $6.95 \times 10^{-2}$ & $9.65 \times 10^{-2}$ & $1.34 \times 10^{-1}$ & $9.45 \times 10^{-2}$ \\
Lamp & $2.20 \times 10^{-5}$ & $9.05 \times 10^{-3}$ & $2.74 \times 10^{-1}$ & $1.56 \times 10^{-1}$ & $6.87 \times 10^{-2}$ & $1.18 \times 10^{-1}$ \\
Chair & $1.09 \times 10^{-3}$ & $1.04 \times 10^{-2}$ & $4.52 \times 10^{-3}$ & $1.05 \times 10^{-1}$ & $9.17 \times 10^{-2}$ & $6.72 \times 10^{-2}$ \\
Chopsticks & $3.64 \times 10^{-3}$ & $1.31 \times 10^{-2}$ & $1.85 \times 10^{-1}$ & $1.87 \times 10^{-1}$ & $1.00 \times 10^{-1}$ & $1.01 \times 10^{-1}$ \\
Cup & $1.59 \times 10^{-3}$ & $5.79 \times 10^{-3}$ & $3.60 \times 10^{-2}$ & $9.84 \times 10^{-2}$ & $6.80 \times 10^{-2}$ & $6.98 \times 10^{-2}$ \\
DeskLamp & $7.62 \times 10^{-4}$ & $8.60 \times 10^{-3}$ & $4.55 \times 10^{-2}$ & $1.30 \times 10^{-1}$ & $7.21 \times 10^{-2}$ & $6.01 \times 10^{-2}$ \\
Dishwasher & $9.66 \times 10^{-3}$ & $2.69 \times 10^{-3}$ & $2.39 \times 10^{-2}$ & $1.27 \times 10^{-1}$ & $4.74 \times 10^{-2}$ & $8.82 \times 10^{-2}$ \\
FloorLamp & $1.23 \times 10^{-4}$ & $2.09 \times 10^{-2}$ & $2.54 \times 10^{-1}$ & $1.68 \times 10^{-1}$ & $4.92 \times 10^{-2}$ & $1.12 \times 10^{-1}$ \\
Fork & $8.81 \times 10^{-3}$ & $2.14 \times 10^{-3}$ & $8.57 \times 10^{-2}$ & $2.14 \times 10^{-1}$ & $1.25 \times 10^{-1}$ & $6.55 \times 10^{-2}$ \\
Hardware & $2.20 \times 10^{-4}$ & $3.07 \times 10^{-3}$ & $4.48 \times 10^{-2}$ & $1.21 \times 10^{-1}$ & $1.02 \times 10^{-1}$ & $1.34 \times 10^{-1}$ \\
Jar & $1.40 \times 10^{-4}$ & $2.44 \times 10^{-3}$ & $6.11 \times 10^{-3}$ & $1.44 \times 10^{-1}$ & $6.31 \times 10^{-2}$ & $8.98 \times 10^{-2}$ \\
LargeShelf & $1.79 \times 10^{-4}$ & $4.65 \times 10^{-3}$ & $5.12 \times 10^{-2}$ & $1.53 \times 10^{-1}$ & $8.67 \times 10^{-2}$ & $7.09 \times 10^{-2}$ \\
Lid & $8.89 \times 10^{-4}$ & $1.09 \times 10^{-2}$ & $1.95 \times 10^{-2}$ & $1.55 \times 10^{-1}$ & $1.22 \times 10^{-1}$ & $1.23 \times 10^{-1}$ \\
LiteDoor & $5.79 \times 10^{-3}$ & $4.39 \times 10^{-3}$ & $2.88 \times 10^{-2}$ & $1.44 \times 10^{-1}$ & $6.32 \times 10^{-2}$ & $9.69 \times 10^{-2}$ \\
LouverDoor & $4.67 \times 10^{-3}$ & $4.84 \times 10^{-3}$ & $9.23 \times 10^{-2}$ & $1.65 \times 10^{-1}$ & $6.82 \times 10^{-2}$ & $1.53 \times 10^{-1}$ \\
Microwave & $3.92 \times 10^{-3}$ & $2.43 \times 10^{-2}$ & $3.15 \times 10^{-2}$ & $7.26 \times 10^{-2}$ & $1.34 \times 10^{-1}$ & $1.65 \times 10^{-1}$ \\
OfficeChair & $1.72 \times 10^{-4}$ & $7.35 \times 10^{-3}$ & $2.95 \times 10^{-2}$ & $8.97 \times 10^{-2}$ & $1.06 \times 10^{-1}$ & $1.05 \times 10^{-1}$ \\
PanelDoor & $9.05 \times 10^{-3}$ & $4.79 \times 10^{-3}$ & $3.74 \times 10^{-2}$ & $1.50 \times 10^{-1}$ & $7.17 \times 10^{-2}$ & $1.09 \times 10^{-1}$ \\
Plate & $1.73 \times 10^{-4}$ & $6.40 \times 10^{-3}$ & $5.78 \times 10^{-3}$ & $1.70 \times 10^{-1}$ & $1.29 \times 10^{-1}$ & $1.74 \times 10^{-1}$ \\
SidetableDesk & $5.11 \times 10^{-5}$ & $3.52 \times 10^{-3}$ & $5.37 \times 10^{-2}$ & $9.64 \times 10^{-2}$ & $5.83 \times 10^{-2}$ & $1.23 \times 10^{-1}$ \\
SimpleBookcase & $3.66 \times 10^{-3}$ & $6.54 \times 10^{-3}$ & $7.01 \times 10^{-3}$ & $1.08 \times 10^{-1}$ & $9.62 \times 10^{-2}$ & $6.06 \times 10^{-2}$ \\
SimpleDesk & $4.29 \times 10^{-5}$ & $9.90 \times 10^{-2}$ & $1.72 \times 10^{-1}$ & $1.68 \times 10^{-1}$ & $6.43 \times 10^{-2}$ & $8.00 \times 10^{-2}$ \\
Sofa & $1.35 \times 10^{-3}$ & $5.78 \times 10^{-3}$ & $6.99 \times 10^{-3}$ & $6.61 \times 10^{-2}$ & $7.32 \times 10^{-2}$ & $6.37 \times 10^{-2}$ \\
Spoon & $5.64 \times 10^{-2}$ & $1.63 \times 10^{-3}$ & $4.59 \times 10^{-2}$ & $2.26 \times 10^{-1}$ & $8.26 \times 10^{-2}$ & $9.81 \times 10^{-2}$ \\
TableCocktail & $2.03 \times 10^{-4}$ & $2.85 \times 10^{-2}$ & $3.10 \times 10^{-2}$ & $1.09 \times 10^{-1}$ & $2.11 \times 10^{-1}$ & $8.68 \times 10^{-2}$ \\
TableDining & $3.31 \times 10^{-3}$ & $7.18 \times 10^{-2}$ & $6.16 \times 10^{-3}$ & $1.55 \times 10^{-1}$ & $1.64 \times 10^{-1}$ & $9.41 \times 10^{-2}$ \\
Toilet & $1.09 \times 10^{-4}$ & $8.44 \times 10^{-3}$ & $1.99 \times 10^{-2}$ & $4.22 \times 10^{-2}$ & $5.24 \times 10^{-2}$ & $5.41 \times 10^{-2}$ \\
TriangleShelf & $3.60 \times 10^{-5}$ & $9.30 \times 10^{-3}$ & $9.47 \times 10^{-2}$ & $1.03 \times 10^{-1}$ & $6.63 \times 10^{-2}$ & $1.08 \times 10^{-1}$ \\
TV & $6.25 \times 10^{-4}$ & $1.74 \times 10^{-3}$ & $1.02 \times 10^{-2}$ & $1.66 \times 10^{-1}$ & $2.84 \times 10^{-2}$ & $6.68 \times 10^{-2}$ \\
TVStand & $2.59 \times 10^{-5}$ & $1.45 \times 10^{-3}$ & $5.63 \times 10^{-2}$ & $1.31 \times 10^{-1}$ & $9.48 \times 10^{-2}$ & $5.92 \times 10^{-2}$ \\
Vase & $9.97 \times 10^{-3}$ & $3.44\times10^{-3}$ & $1.05\times10^{-1}$ & $2.68 \times 10^{-1}$ & $2.48\times10^{-2}$ & $4.00\times10^{-2}$ \\
Window & $9.57 \times 10^{-3}$ & $3.51 \times 10^{-3}$ & $6.61 \times 10^{-2}$ & $1.81 \times 10^{-1}$ & $1.14 \times 10^{-1}$ & $1.88 \times 10^{-1}$ \\
Wineglass & $1.40 \times 10^{-2}$ & $3.12 \times 10^{-3}$ & $3.86 \times 10^{-2}$ & $1.04 \times 10^{-1}$ & $5.39 \times 10^{-2}$ & $6.99 \times 10^{-2}$ \\
\midrule
\textbf{All (Std.)} & $\mathbf{2.92 \times 10^{-3}}$ & $\mathbf{2.49 \times 10^{-2}}$ & $\mathbf{7.23 \times 10^{-2}}$ & $\mathbf{1.25 \times 10^{-1}}$ & $\mathbf{1.94 \times 10^{-1}}$ & $\mathbf{1.92 \times 10^{-1}}$ \\
\bottomrule
\end{tabular}
}
\end{table*}

In addition to evaluating our object-to-code inference model, we also perform a quantitative assessment of our part-to-code inference model. Specifically, for each category described in Section~\ref{sec: part_and_code_dataset}, we construct a test set consisting of 10{,}000 samples. We evaluate the model's performance using the CD and voxel IoU metrics on these test sets. The results, shown in Table~\ref{tab:part_results}, demonstrate strong performance across all categories, with low CD values and high IoU scores, indicating that our part-to-code inference model is highly effective in generating accurate code representations for individual parts.

\begin{table}[h]
\centering
\caption{Quantitative evaluation of the \textit{part-to-code} inference model across different part categories. CD is reported in $10^{-2}$, and IoU is reported in percentage.}
\label{tab:part_results}
\begin{tabular}{lcc}
\toprule
\textbf{Category} & \textbf{CD ($\times 10^{-2}$)} & \textbf{IoU (\%)} \\
\midrule
Primitive     & 0.18  & 94.81 \\
Boolean          & 0.03  & 96.13 \\
Array         & 0.70  & 78.90 \\
Bridge Loop  & 0.14  & 89.16 \\
Translation   & 0.17  & 83.45 \\
\bottomrule
\end{tabular}
\end{table}

\subsection{Complete experiment result of Shape Editing}
\label{appendix: Shape Editing}
\begin{figure}[ht]
    \centering
    \includegraphics[width=14cm]{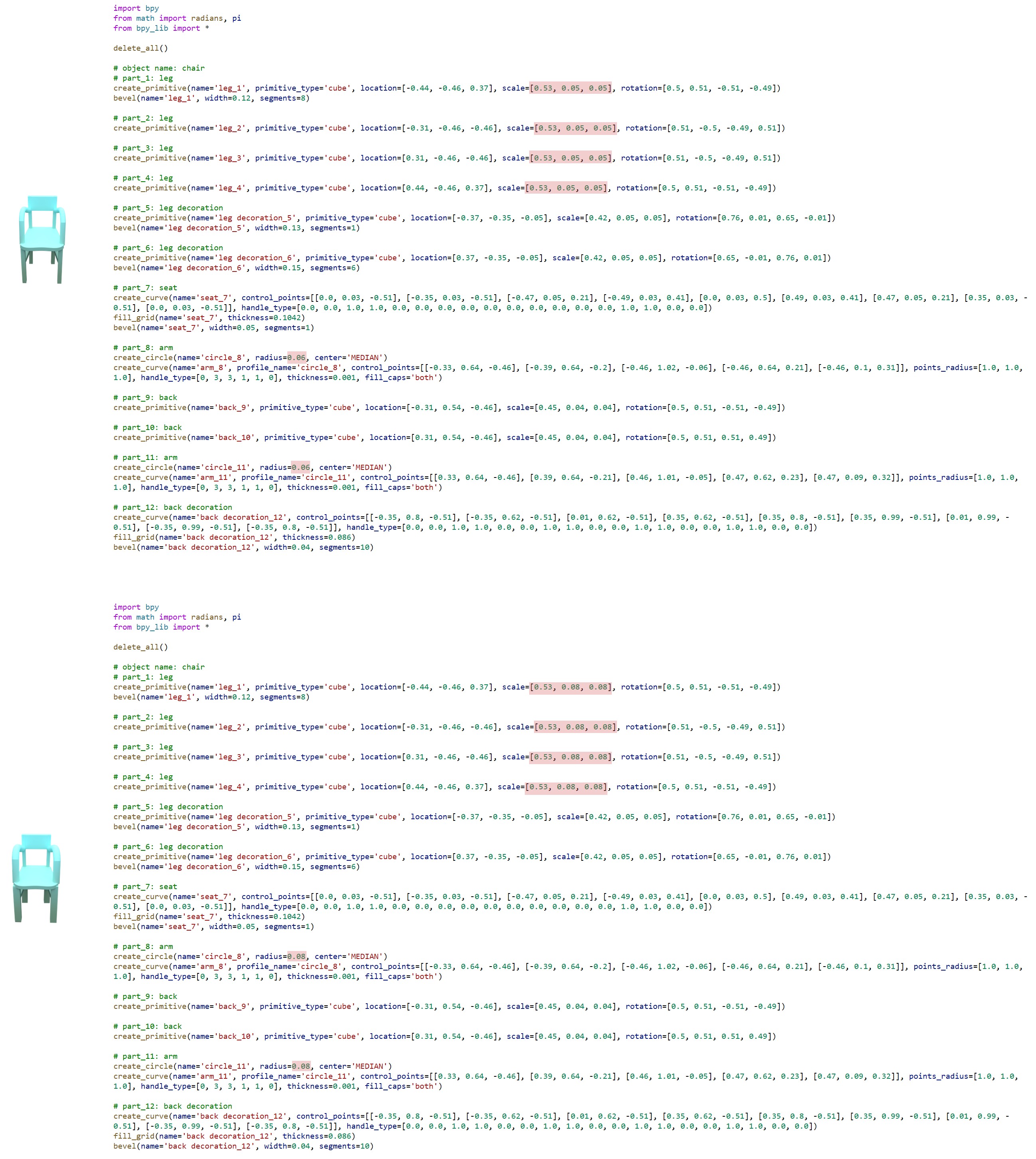}
    \caption{
    By modifying the \texttt{scale} parameters of the \texttt{leg} and \texttt{arm} parts, we adjust their thickness. The highlighted sections indicate the changes made.
}
    \label{fig:edit-chair.}
\end{figure}

\begin{figure}[ht]
    \centering
    \includegraphics[width=14cm]{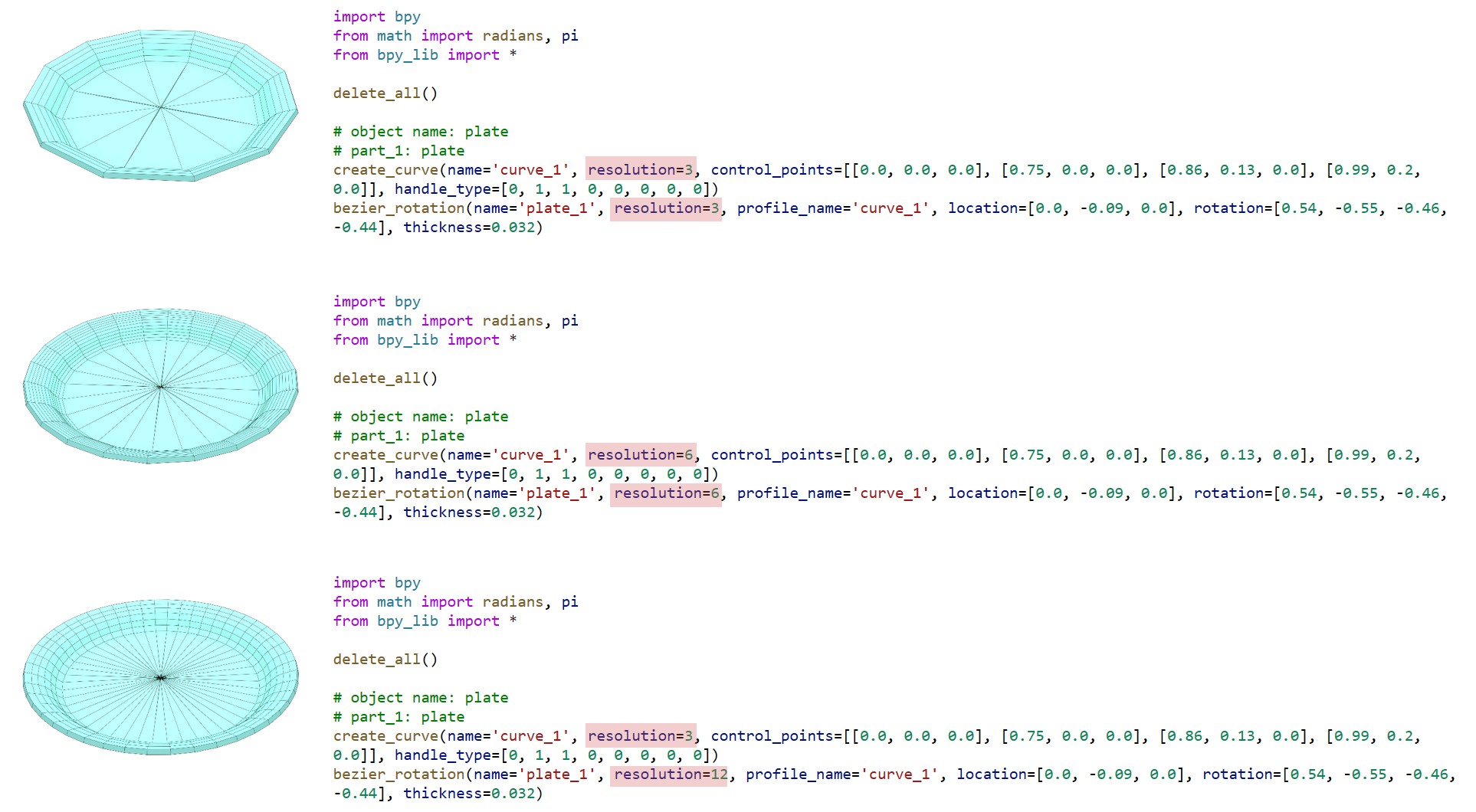}
    \caption{
    By modifying the \texttt{resolution} parameter, we change its resolution. The highlighted sections indicate the changes made.}
    \label{fig:edit-plate.}
\end{figure}

We additionally present two examples of shape editing along with their complete code implementations. In Figure~\ref{fig:edit-chair.}, we modify the thickness of the chair legs and armrests by adjusting the \texttt{scale} parameter. In Figure~\ref{fig:edit-plate.}, we change the mesh resolution of a plate by modifying the \texttt{resolution} parameter.

\subsection{Complete experiment result of Shape Understanding}
\label{appendix: Shape Undertanding}

\begin{figure}[ht]
    \centering
    \includegraphics[width=14cm]{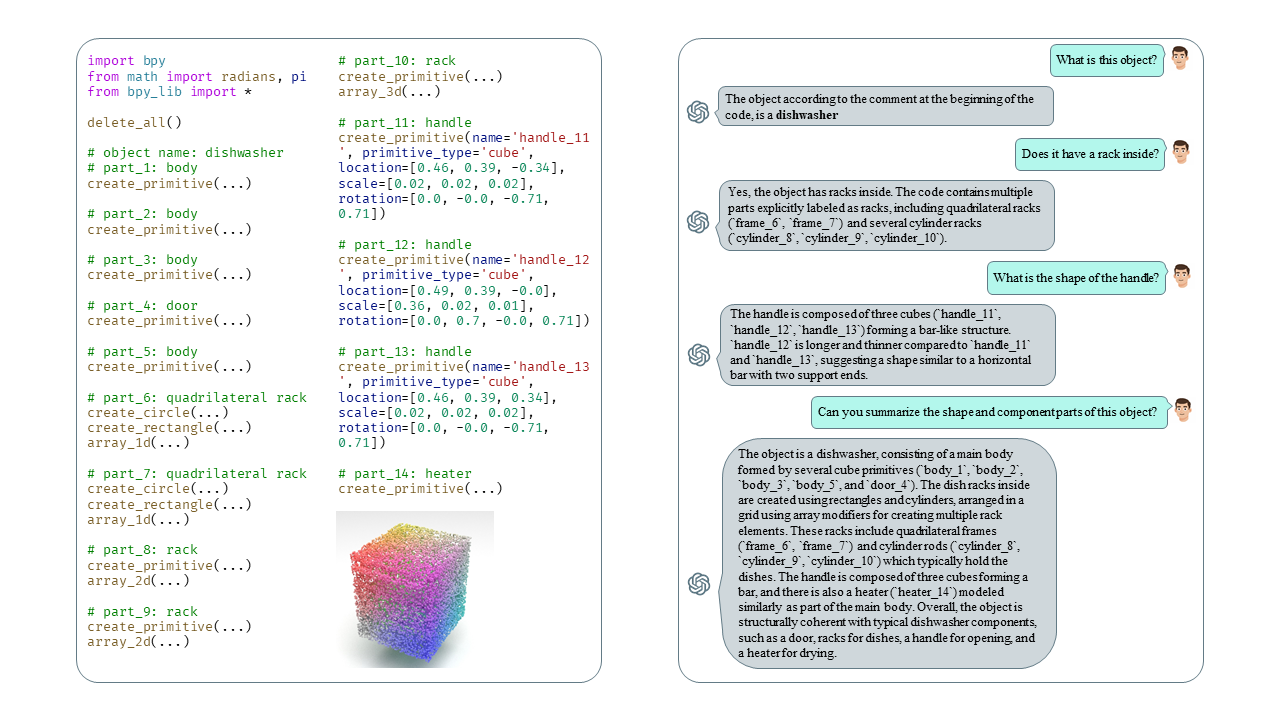}
    \caption{
    Experiments on how GPT-4o can understand shape through given dishwasher code.}
    \label{fig:dish}
\end{figure}

\begin{figure}[ht]
    \centering
    \includegraphics[width=14cm]{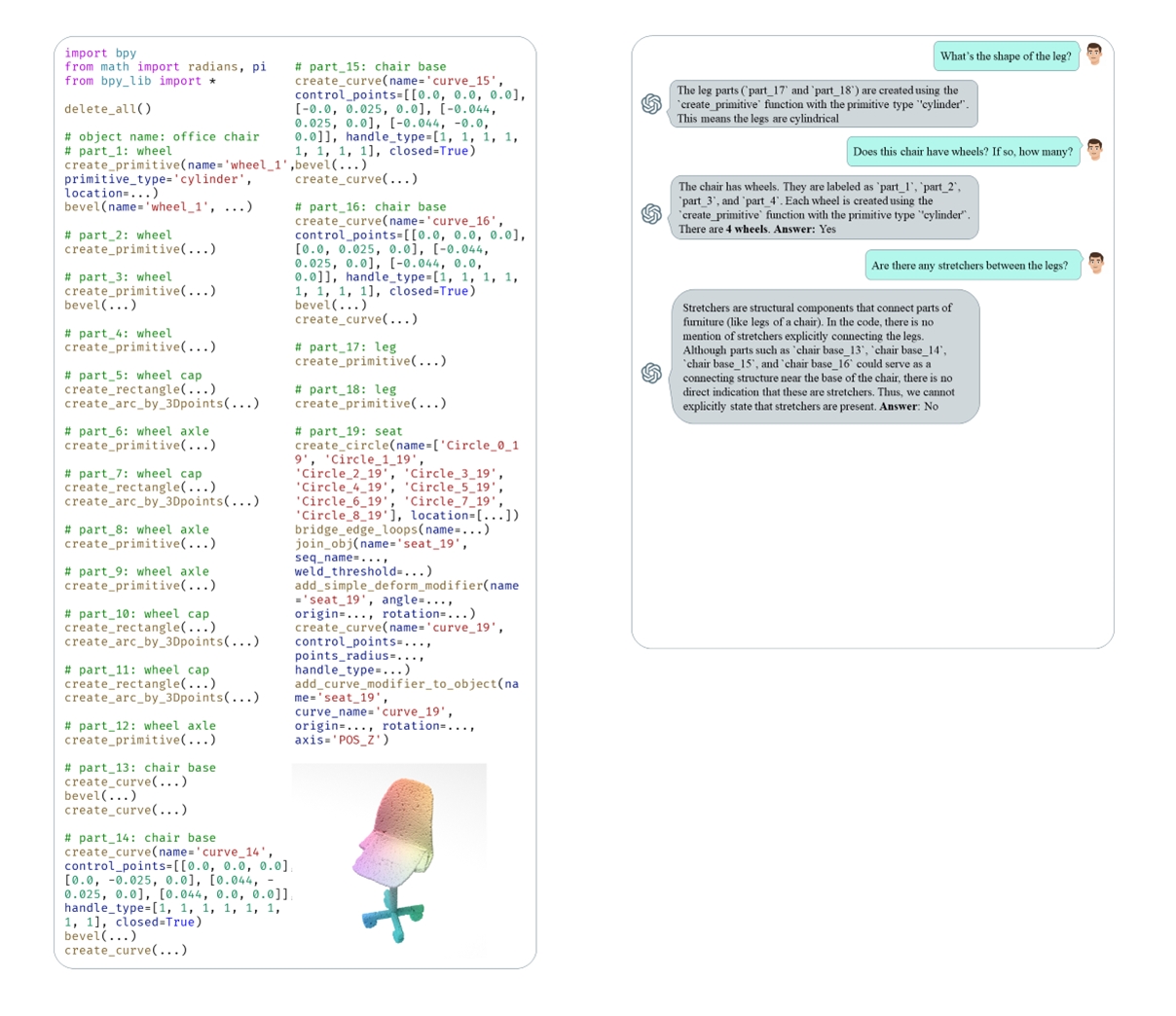}
    \caption{
    Experiments on how GPT-4o can understand shape through given office chair code.}
    \label{fig:off}
\end{figure}

\begin{figure}[ht]
    \centering
    \includegraphics[width=14cm]{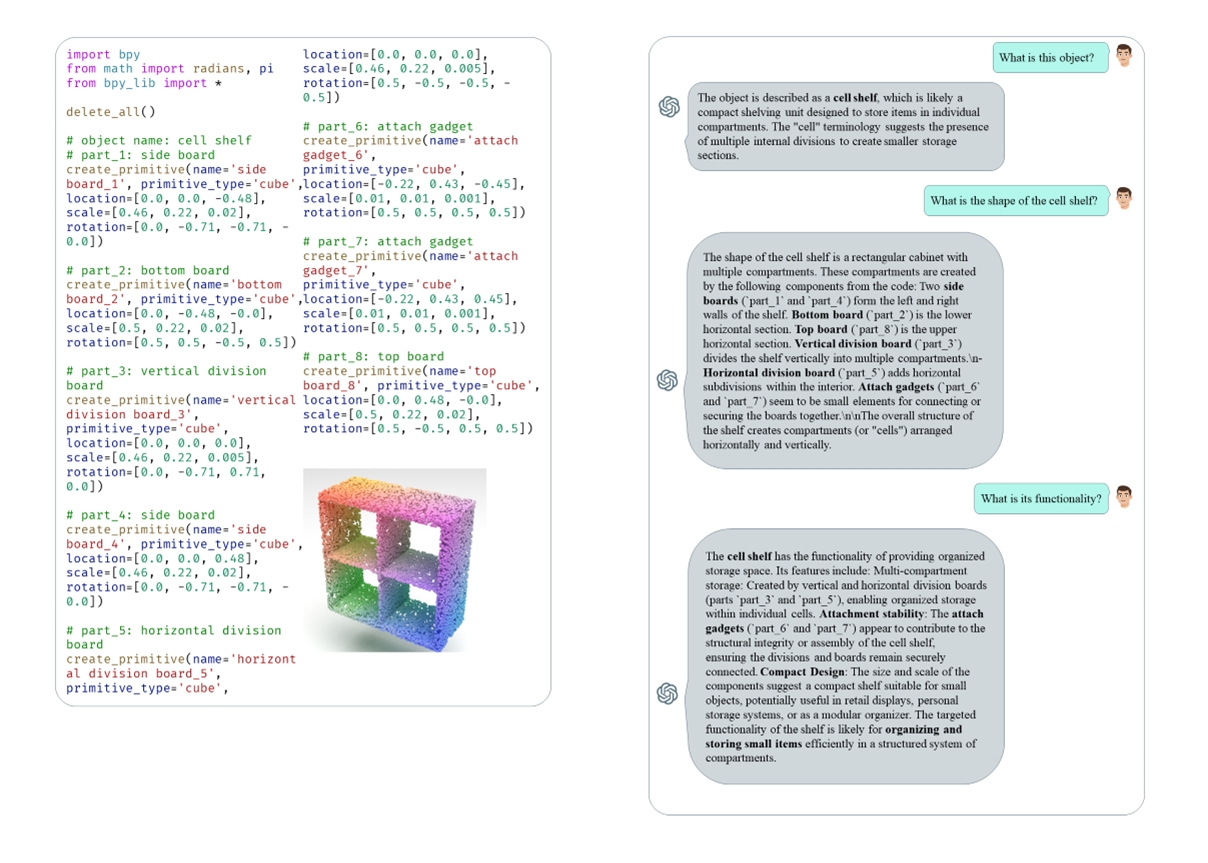}
    \caption{
    Experiments on how GPT-4o can understand shape through given cell shelf code.}
    \label{fig:cell}
\end{figure}

When presented with a 3D point cloud of an object as input,  MeshCoder can infer the corresponding code for the object. Upon execution of this code in Blender, the geometry of the object can be obtained. Notably, the comments within the code encompass a variety of semantically rich cues, such as the object's identity and the specifics of each component. The primary aim of this experiment is to highlight that our model can assist existing large language models, like GPT - 4, in understanding the structure of 3D objects. We provide the inferred code to GPT - 4 and then inquire about the geometry or structure of the object, as showed in Figure~\ref{fig:dish}, Figure~\ref{fig:off} and Figure~\ref{fig:cell}. GPT - 4 is able to generate relevant responses based on the code inferred by our model. This demonstrates that our model possesses capabilities in understanding the geometry and structure of 3D objects and can aid large - scale models such as GPT in addressing such questions. However, our model does have limitations. Currently, the code inferred by our model solely contains geometric information of the object and does not include color information. As a result, it is unable to answer questions pertaining to color.

\clearpage

\clearpage
\begin{figure}[ht]
    \vspace{-2cm}
    \centering
    \includegraphics[width=14cm]{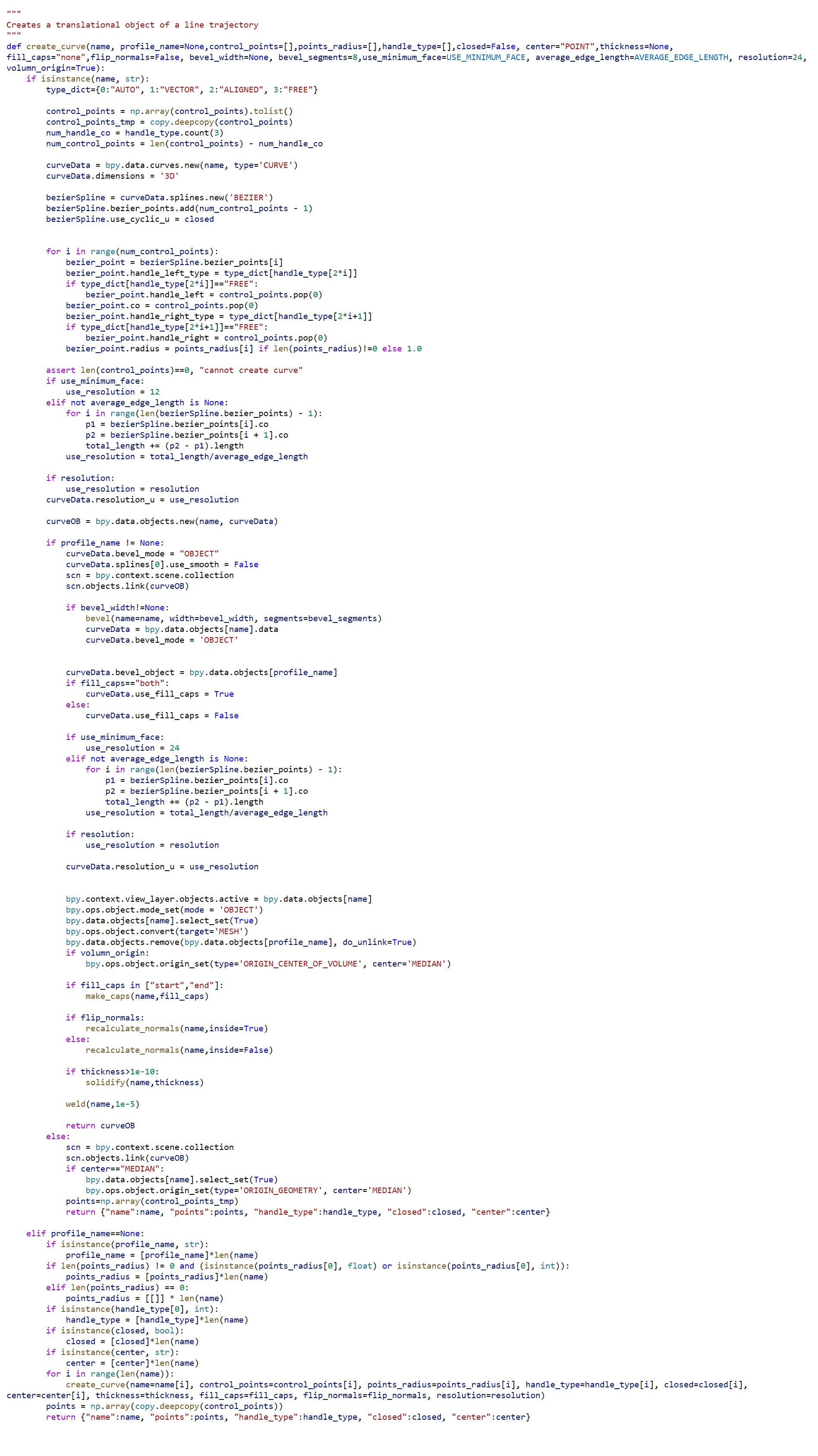}
    \vspace{-0.5cm}
    \caption{
    Implementation of the function for creating curves}
    \label{fig:code_curve}
\end{figure}

\end{document}